\documentclass[a4paper,11pt]{article}
\pdfoutput=1 
\usepackage{jheparxiv}
\usepackage[T1]{fontenc}
\usepackage[normalem]{ulem}

\usepackage{tensor}
\usepackage{amsmath}
\usepackage{amssymb}
\usepackage{mathrsfs}
\usepackage{dsfont}
\usepackage{graphicx}
\usepackage{stmaryrd}
\usepackage{twistor}
\usepackage{tikz}
\usepackage{commath}
\usepackage{xcolor}
\usepackage{mathtools}
\usepackage{tkz-euclide,subfigure}
\usepackage{capt-of}
\usepackage{tikz-cd}
\usepackage{comment}

\newcommand{\veps}{\varepsilon}

\newcommand{\sa}{\mathsf{a}}
\newcommand{\nbar}{\bar{\nabla}}

\newcommand{\msf}[1]{\mathsf{#1}}
\newcommand{\mbf}[1]{\mathbf{#1}}
\newcommand{\mbb}[1]{\mathbb{#1}}
\newcommand{\mfk}[1]{\mathfrak{#1}}
\newcommand{\mcal}[1]{\mathcal{#1}}

\newcommand{\dt}[1]{\dot{#1}}
\newcommand{\bv}[1]{\Breve{#1}}

\title{Twistorial chiral algebras in higher dimensions}

\author{Tim Adamo} 
\author{\& Iustin Surubaru}

\affiliation{School of Mathematics and Maxwell Institute for Mathematical Sciences \\
        University of Edinburgh, EH9 3FD, United Kingdom}

\emailAdd{t.adamo@ed.ac.uk}
\emailAdd{iustin.surubaru@ed.ac.uk}

\abstract{In four spacetime dimensions, the classically integrable self-dual sectors of gauge theory and gravity have associated chiral algebras, which emerge naturally from their description in twistor space. We show that there are similar chiral algebras associated to integrable sectors of gauge theory and gravity whenever the spacetime dimension is an integer multiple of four. In particular, the hyperk\"ahler sector of gravity and the hyperholomorphic sector of gauge theory in $4m$-dimensions have well-known twistor descriptions giving rise to chiral algebras. Using twistor sigma models to describe these sectors, we demonstrate that the chiral algebras in higher-dimensions also arise as soft symmetry algebras under a certain notion of collinear limit. Interestingly, the chiral algebras and collinear limits in higher-dimensions are defined on the 2-sphere, rather than the full celestial sphere.}

\begin{document}
\maketitle
\flushbottom

\section{Introduction}

In quantum field theory (QFT), the external states of scattering processes are typically represented in a momentum eigenstate basis, but alternative bases of solutions to the free field equations can shed light on otherwise hidden aspects of the resulting scattering amplitudes. A particularly fruitful example is the \emph{conformal primary basis}~\cite{Pasterski:2016qvg,Pasterski:2017kqt}, in which free fields are parametrized by a (complex) scaling dimension and a point on the celestial sphere of null directions. In $d$ spacetime dimensions, these conformal primary states transform as conformal primaries on the $(d-2)$-dimensional celestial sphere under the action of the Lorentz group, and scattering amplitudes computed in this basis -- often called \emph{celestial amplitudes} -- behave like conformal correlators on the celestial sphere~\cite{Pasterski:2016qvg,Pasterski:2017kqt,Pasterski:2017}.

The study of celestial amplitudes and the quest to construct conformal field theories (CFTs) on the celestial sphere which produce them dynamically is now referred to as `celestial holography' (cf., \cite{Raclariu:2021zjz,Pasterski:2021rjz,Pasterski:2021raf,McLoughlin:2022ljp,Pasterski:2023ikd,Donnay:2023mrd} for recent reviews). One of the most interesting outputs of celestial holography has been the realization that graviton and gluon scattering amplitudes in four-dimensional Minkowski spacetime encode \emph{chiral algebras} associated with each helicity sector: these are infinite-dimensional meromorphic vertex operator algebras defined on the two-dimensional celestial sphere. These chiral algebras are encoded in the infrared structures of 4d celestial graviton and gluon amplitudes. The infinite tower of `conformally soft' gluons and gravitons of positive helicity -- given by certain integer values of the scaling dimension which render the conformal primary wavefunction singular~\cite{Donnay:2018neh,Banerjee:2019aoy,Fan:2019emx,Pate:2019mfs,Nandan:2019jas,Adamo:2019ipt,Puhm:2019zbl} and correspond to a soft expansion in momentum space~\cite{Guevara:2019ypd} -- form a closed algebra under holomorphic collinear limits on the celestial 2-sphere~\cite{Guevara:2021abz}. These holomorphic collinear limits are equivalent to holomorphic OPE limits from the perspective of the 2-sphere~\cite{Fan:2019emx,Pate:2019lpp}, so the resulting algebra automatically has the structure of a chiral algebra.

A non-trivial re-definition of the modes shows that these chiral algebras are simply $\cL\mathfrak{ham}(\C^2)$ for positive helicity gravitons and $\cL\mathfrak{g}[\C^2]$ for positive helicity gluons (where $\mathfrak{g}$ is the Lie algebra of the gauge group)~\cite{Strominger:2021mtt}. These are the loop algebras of the Lie algebras of Hamiltonian vector fields on $\C^2$ and polynomial maps from $\C^2$ into $\mathfrak{g}$, respectively. In the gravitational case, $\cL\mathfrak{ham}(\C^2)$ is closely related to $\cL w^{\wedge}_{1+\infty}$~\cite{Hoppe:1988gk,Bakas:1989xu,Bakas:1989mz}, the loop algebra of the wedge algebra of $w_{1+\infty}$, and it is sometimes referred to as such.   

While the appearance of these chiral algebras from an amplitudes perspective is surprising, it follows straightforwardly from the perspective of \emph{twistor theory}~\cite{Penrose:1967wn}, a mathematical framework which trivializes the self-dual sectors of gauge theory and gravity, manifesting their classical integrability~\cite{Ward:1977ta,Penrose:1976js}. In particular, the conformally soft tower of positive helicity gravitons or gluons is realized as the symmetries of the self-dual sector in twistor space, which form the chiral algebras under the natural twistorial structures~\cite{Adamo:2021lrv}\footnote{Indeed, the connection between self-duality and infinite-dimensional symmetry algebras via twistor theory had been realized in different guises by many others over the years~\cite{Boyer:1985aj,Park:1989fz,Park:1989vq,Mason:1990,Dunajski:2000iq}.}. These chiral algebras have now been studied extensively: their appearance has been verified at tree-level in generic effective field theories, extends to an action on hard (rather than conformally soft) states and can be interpreted in terms of asymptotic symmetries~\cite{Himwich:2021dau,Jiang:2021ovh,Mago:2021wje,Ren:2022sws,Himwich:2023njb,Agrawal:2024sju}; they can be understood from a covariant phase space perspective~\cite{Freidel:2021dfs,Freidel:2021ytz,Freidel:2023gue,Mason:2023mti,Donnay:2024qwq,Geiller:2024bgf,Kmec:2024nmu}; and they have been generalised to settings with background curvature~\cite{Costello:2022jpg,Garner:2023izn,Costello:2023hmi,Bittleston:2023bzp,Adamo:2023zeh,Taylor:2023ajd,Bittleston:2024rqe,Adamo:2024xpc,Bogna:2024gnt}, non-commutativity~\cite{Monteiro:2022lwm,Bu:2022iak}, higher-spin fields~\cite{Monteiro:2022xwq,Ponomarev:2022ryp} and supersymmetry~\cite{Fotopoulos:2020bqj,Brandhuber:2021nez,Jiang:2021xzy,Bu:2021avc,Ahn:2022oor,Ahn:2024kpv,Crawley:2024cak,Tropper:2024evi}. 

The presence of infinite-dimensional chiral algebras in the self-dual sectors of gauge theory and gravity places an enormous amount of constraints on scattering amplitudes, particularly for helicity configurations `close' to self-duality, such as maximal-helicity-violating (MHV) amplitudes. For instance, the chiral algebras are linked with the existence of all-order collinear expansions for MHV amplitudes~\cite{Adamo:2022wjo,Ren:2023trv} and underpin the only known top-down constructions of celestial holography~\cite{Costello:2022jpg,Costello:2023hmi,Zeng:2023qqp,Bittleston:2024efo}. Furthermore, chiral algebras enable bootstrap constructions of certain scattering amplitudes to high loop-orders~\cite{Costello:2023vyy,Dixon:2024mzh,Dixon:2024tsb} or in non-trivial backgrounds~\cite{Garner:2024tis}.

\medskip

Given the power of celestial chiral algebras in 4d, it seems natural to ask if similar infinite-dimensional algebras are present in gravity or gauge theory in higher spacetime dimensions. In particular, these chiral algebras are closely linked with remarkable scattering amplitude formulae (e.g., the Parke-Taylor formula for MHV gluon scattering at tree level~\cite{Parke:1986gb}), all known examples of which are confined to $d\leq4$ spacetime dimensions. Finding celestial chiral algebras in higher spacetime dimensions would thus be a smoking gun indicating the existence of similarly remarkable amplitude formulae in higher-dimensions, which in turn would provide a platform for the pursuit of celestial holography in more general settings. 

At first, this seems unlikely, though: helicity states for gravitons and gluons, and their non-linear extension to self-dual sectors, are special to $d=4$, and in general $d>4$ there are no integrable subsectors of general relativity or Yang-Mills theory. Furthermore, many of the celestial structures present in 4d are absent in higher dimensions. Indeed, the global conformal group on the celestial sphere $S^{d-2}$ does not receive an infinite-dimensional enhancement for $d>4$ and the asymptotic symmetry group can be either finite or infinite-dimensional, depending on the definition of asymptotic flatness used (cf., \cite{Hollands:2003ie,Tanabe:2011es,Hollands:2016oma,Mao:2017wvx,Pate:2017fgt,Aggarwal:2018ilg,He:2019pll,Henneaux:2019yqq,Campoleoni:2019ptc,Campoleoni:2020ejn,Capone:2021ouo}). Additionally, the soft symmetry algebras in $d>4$ have been shown to be finite-dimensional in general~\cite{Pano:2023slc}.

Nevertheless, in this paper we uncover infinite-dimensional symmetry algebras in both gauge theory and gravity whenever the spacetime dimension is a multiple of four: $d=4m$. The key idea is to use twistor theory, which gives such a natural description of chiral symmetry algebras of self-dual sectors in 4d, in higher-dimensions. In $d=4m$, there are well-known twistor constructions of \emph{hyperk\"ahler} metrics~\cite{Salamon:1982,Hitchin:1986ea} and \emph{hyperholomorphic} gauge fields~\cite{Ward:1983zm}: roughly speaking, these are field configurations compatible with a 2-sphere's worth of integrable complex structures, with the associated twistor construction being the `total space' of these complex structures over spacetime. 

Much like the self-dual sectors in 4d, hyperk\"ahler metrics and hyperholomorphic gauge fields are classically integrable subsectors of vacuum general relativity and Yang-Mills theory in $4m$-dimensions. Exploiting their twistor descriptions, it is straightforward to determine the symmetry algebras of hyperk\"ahler gravitational perturbations and hyperholomorphic gauge perturbations around flat space; remarkably, these are again chiral algebras (now $\cL\mathfrak{ham}(\C^{2m})$ and $\cL\mathfrak{g}[\C^{2m}]$), defined on the \emph{two-dimensional} Riemann spheres in twistor space corresponding to points in spacetime. These sectors also have dynamical descriptions in terms of chiral 2d CFTs known as `twistor sigma models'~\cite{Witten:2003nn,Berkovits:2004hg,Adamo:2021bej}, with the chiral algebras emerging as an infinite-dimensional charge algebra (under semiclassical OPE).  

One might wonder what, if any, connection these chiral algebras could have with celestial holography when the celestial sphere is $(d-2)$-dimensional. We show that hyperk\"ahler gravitons and hyperholomorphic gluons have on-shell (complex) momenta which are \emph{non-generic}: in particular, they are parameterized by a frequency and a point on $\P^1\times\P^{2m-1}$ inside of the complexified celestial sphere. Consequently, there is a notion of collinear limit between hyperk\"ahler gravitons or hyperholomorphic gluons which is entirely controlled by the $\P^1$ factor and therefore two-dimensional in nature. We show that the chiral algebras emerge as soft symmetry algebras under this two-dimensional collinear limit, and derive the action of these chiral algebras on `hard' hyperk\"ahler or hyperholomorphic states.  

\medskip

The paper is organized as follows. In Section~\ref{sec 2}, we review the higher-dimensional hyperk\"ahler sector of gravity and hyperholomorphic sector of gauge theory, along with their twistor descriptions. Section~\ref{sec 3} derives the symmetry algebras of these sectors around the trivial vacuum using the geometric structures of twistor space. In Section~\ref{sec 4}, we review the twistor sigma models for the hyperk\"ahler and hyperholomorphic sectors and show that they give rise to the twistorial chiral algebras for these sectors through their OPEs. Section~\ref{sec 5} explores the kinematics of hyperk\"ahler gravitons and hyperholomorphic gluons and defines their conformally soft wavefunctions, which can be expanded in modes of the chiral algebras. We then show that the action of the chiral algebras on hard wavefunctions takes the form of holomorphic OPE. Section~\ref{sec 6} concludes.


\section{Integrable sectors of gauge theory and gravity in higher-dimensions}
\label{sec 2}

The study of twistorial, or celestial, chiral algebras in four spacetime dimensions is closely tied to the existence of a consistent, classically integrable subsector of gauge theory and gravity in 4d. Indeed, the \emph{self-dual sectors} of gauge theory and gravity -- defined by the vanishing of the anti-self-dual part of the field strength and Weyl tensor, respectively -- automatically solve the vacuum Yang-Mills and Einstein equations, and are classically integrable. In the interpretation of chiral algebras as soft symmetry algebras, this is manifest as the algebras are defined under (holomorphic) collinear limits among gluons or gravitons of the same helicity~\cite{Fan:2019emx,Pate:2019lpp,Guevara:2021abz,Strominger:2021mtt}. Furthermore, the chiral algebras emerge naturally from the twistor constructions of self-dual gauge theory and gravity, which make the integrability of these sectors explicit~\cite{Adamo:2021lrv}.

At first glance, it seems hopeless to look for such structures in gauge theory or gravity in spacetime dimensions greater than four. Self-duality is clearly very special to four-dimensions, being intimately related to the fact that the Hodge star acts involutively on the space of 2-forms, where the gauge invariant field strengths live. While it is certainly true that there is generically no analogy of the self-dual sector for spacetime dimension $d>4$, there \emph{are} natural generalisations when $d=4m$, for $m$ any positive integer. These are known as the \emph{hyperk\"ahler} and \emph{hyperholomorphic} sectors of gravity and gauge theory, respectively. They have many of the features of the self-dual sector when $m=1$: they are solutions to the vacuum Einstein and Yang-Mills equations, are classically integrable and admit descriptions in terms of twistor theory -- all of the ingredients needed to give rise to infinite-dimensional symmetry algebras.

In this section, we review the definitions of hyperk\"ahler geometry and hyperholomorphic gauge fields as well as their twistor constructions. The reader who is already acquainted with these concepts can safely skim this material to familiarize themselves with our notation.


\subsection{Hyperk\"ahler metrics \& their twistor theory}

A $4m$-dimensional Riemannian manifold is said to be \emph{hyperk\"ahler} if the holonomy group of its metric is contained in Sp$(m)$. This special holonomy implies that all hyperk\"ahler manifolds are Ricci flat, and thus solutions to the vacuum Einstein equations (see Chapter 14 of~\cite{Besse:1987}, Chapter 13 of~\cite{Berger:2003}, Chapter 10 of~\cite{Joyce:2007} and~\cite{Verbitsky:2010} for reviews). Equivalently, a hyperk\"ahler manifold admits three integrable complex structures, each of which is K\"ahler with respect to the metric, and which form the quaternionic algebra. This is in turn equivalent to the statement that there exists a 2-sphere's worth of integrable complex structures at each point on the manifold. In 4-dimensions (i.e., $m=1$), the hyperk\"ahler condition is equivalent to the vacuum self-duality equations.

To study massless field theories in this context, it is natural to consider \emph{complexified hyperk\"ahler} manifolds, which have complex dimension $4m$ and are endowed with a holomorphic metric whose holomomy is contained in Sp$(m,\C)$. While Riemannian signature (as well as indefinite signature) real slices can be obtained by specifying reality conditions, it will be this complexified setting that we consider from now on\footnote{Note that complex hyperk\"ahler structures also arise naturally in the study of stability conditions associated with Donaldson-Thomas theory~\cite{Bridgeland:2019fbi,Bridgeland:2020zjh,Alexandrov:2021wxu}.}, and we will abuse terminology by using `hyperk\"ahler' (HK) rather than `complex hyperk\"ahler' in what follows.  

The simplest example of a HK manifold is the flat model given by $\C^{4m}$. Let $x^{\alpha\dot\alpha}$ be holomorphic coordinates on $\C^{4m}$, where $\alpha=0,1$ is a SL$(2,\C)$ spinor index and $\dot\alpha=1,\ldots,2m$ is a Sp$(m,\C)$ index. The flat metric can then be written in a way that manifests the decomposition of the tangent bundle into the tensor product of (flat) SL$(2,\C)$ and Sp$(m,\C)$ bundles (a special case of a paraconformal structure~\cite{Bailey:1991}):
\be\label{flatmet}
\d s^2_{\C^{4m}}=\veps_{\alpha\beta}\,\veps_{\dot\alpha\dot\beta}\,\d x^{\alpha\dot\alpha}\,\d x^{\beta\dot\beta}\,,
\ee
where $\veps_{\alpha\beta}$ is the SL$(2,\C)$-invariant Levi-Civita symbol and $\veps_{\dot\alpha\dot\beta}$ is the symplectic form of Sp$(m,\C)$:
\be\label{sympforms}
\veps_{\alpha\beta}=\left(\begin{array}{cc}
                          0 & 1 \\
                          -1 & 0
                          \end{array}\right)\,, \qquad 
\veps_{\dot\alpha\dot\beta}=\left(\begin{array}{cc}
                                  0_{m} & \mathds{1}_{m} \\
                                  -\mathds{1}_{m} & 0_{m}
                                  \end{array}\right)\,,
\ee
for $0_m$ the $m\times m$ zero matrix and $\mathds{1}_m$ the $m\times m$ identity matrix. These objects (and their inverses) can be used to raise and lower spinor indices according to the conventions
\be\label{spinorconvention}
a^{\alpha}\,\veps_{\alpha\beta}=a_{\beta}\,, \qquad \veps^{\alpha\beta}\,a_\beta=a^{\alpha}\,,
\ee
and similarly for dotted indices. It will often be useful to introduce some shorthand for SL$(2,\C)$ and Sp$(m,\C)$ invariant inner products:
\be\label{asqbrackets}
\la a\,b\ra:=a^{\alpha}\,b_{\alpha}\,, \qquad [\tilde{a}\,\tilde{b}]:=\tilde{a}^{\dot\alpha}\,\tilde{b}_{\dot\alpha}\,,
\ee
respectively. Note that both of these inner products are skew-symmetric in their arguments.

Of course, the holonomy of the flat metric \eqref{flatmet} is trivially contained in Sp$(m,\C)$; the advantage of working with spinors to manifest the paraconformal structure is more clear when looking at the hyperk\"ahler structure in terms of complex structures. Consider the triplet of closed 2-forms:
\be\label{flat2forms}
\Sigma^{\alpha\beta}=\d x^{\alpha\dot\alpha}\wedge\d x^{\beta}{}_{\dot\beta}\,,
\ee
which are automatically closed $\d\Sigma^{\alpha\beta}=0$. These in turn define a triplet of (complex) K\"ahler forms for the metric
\be\label{Ktriple}
\Sigma^{01}\,, \qquad \Sigma^{00}\pm\im\,\Sigma^{11}\,,
\ee
whose associated complex structures are easily seen to be integrable and form the quaternion algebra. Hence, there is a sphere bundle of complex structures over $\C^{4m}$; the total space of this bundle is the \emph{twistor space} of $\C^{4m}$. 

More precisely, let $Z^{A}=(\mu^{\dot\alpha},\lambda_{\alpha})$ be holomorphic, homogeneous coordinates on the complex projective space $\P^{2m+1}$ and define the twistor space
\be\label{twistorflat}
\PT=\left\{[Z]\in\P^{2m+1}\,|\,\lambda_{\alpha}\neq 0\right\}\,,
\ee
where $[Z]$ indicates the projective equivalence class $Z^{A}\sim r\,Z^A$ of the homogeneous coordinates, for $r$ any non-vanishing complex number. In other words, twistor space is the open subset of $\P^{2m+1}$ obtained by removing the $\P^{2m-1}$ corresponding to $\lambda_{\alpha}=0$. As such, $\PT$ admits a holomorphic fibration over the Riemann sphere $\pi:\PT\to\P^1$, with $\lambda_{\alpha}$ providing the holomorphic homogeneous coordinates on the base. The $\C^{2m}$ fibres of this fibration are equipped with a weighted, holomorphic Poisson structure defined by
\be\label{Pstruct}
    I:=\veps^{\dal\dt{\beta}}\,\frac{\del}{\del\mu^{\dal}}\wedge \frac{\del}{\del\mu^{\dt{\beta}}}\,,
\ee
taking values in $\cO(-2)$.

Each point $x\in\C^{4m}$ now corresponds to a global section of the fibration $\pi$ given by
\be\label{incidence}
\mu^{\dot\alpha}=x^{\alpha\dot\alpha}\,\lambda_{\alpha}\,.
\ee
In other words, points in $\C^{4m}$ are given by linear, holomorphic Riemann spheres in $\PT$, often referred to as `twistor lines.' The relation between $\PT$ and the HK structure on $\C^{4m}$ is clear: \eqref{incidence} arranges the coordinates $x^{\alpha\dot\alpha}$ in a sphere's worth (encoded by $\lambda_\alpha$) of $\mu^{\dot\alpha}$ which are holomorphic with respect to the corresponding complex structures of the HK structure. Indeed, the triplet of 2-forms \eqref{flat2forms} encoding the HK structure is recovered on twistor space by taking
\be\label{Gindikin}
\d_{x}\mu^{\dot\alpha}\wedge\d_{x}\mu_{\dot\alpha}=\Sigma^{\alpha\beta}\,\lambda_{\alpha}\,\lambda_{\beta}\,,
\ee
where $\d_x\mu^{\dot\alpha}$ denotes the exterior derivative with respect to $x$ of $\mu^{\dot\alpha}$ evaluated on the global section \eqref{incidence}.

\medskip

Remarkably, it can be shown that \emph{every} HK manifold has a twistor space, and that every twistor space with the basic structures of $\PT$ (i.e., an integrable complex structure, holomorphic fibration over $\P^1$ and holomorphic Poisson structure on the fibres) corresponds to a HK manifold. This was first established in 4-dimensions ($m=1$) with the famous non-linear graviton theorem of Penrose~\cite{Penrose:1976js}, and subsequently extended to general $m>1$ HK manifolds~\cite{Atiyah:1978wi,Salamon:1982,Hitchin:1986ea,LeBrun:1989}. The precise statement is:
\begin{thm}\label{Thm:NLGrav}
    There is a one-to-one correspondence between:
    \begin{itemize}
        \item suitably convex regions of hyperk\"ahler $4m$-manifolds, and
        \item $(2m+1)$-dimensional complex manifolds $\CPT$ that are complex deformations of a neighbourhood of a twistor line in $\PT$, preserving the holomorphic fibration $\pi:\CPT\to\P^1$ and weighted Poisson structure $I$.
    \end{itemize}
\end{thm}

In practical terms, this theorem states that any HK manifold can be obtained from a complex deformation of the twistor space of the flat model. If $\dbar$ denotes the anti-holomorphic Dolbeault operator for the natural complex structure on $\P^{2m+1}$, then such a deformation can be expressed in terms of a deformed Dolbeault operator
\be\label{defDol}
\nbar=\dbar+V\,, \qquad V\in\Omega^{0,1}(\PT, T_{\PT})\,,
\ee
where $T_{\PT}$ is the holomorphic tangent bundle of $\PT$. The requirements that this deformation preserve the holomorphic fibration over $\P^1$ and the weighted Poisson structure on its fibres mean that $V$ must actually be Hamiltonian with respect to the (weighted) symplectic structure induced on the fibres of $\pi$ by $I$. In other words, the deformation must take the form
\be\label{defDol2}
V=\frac{\partial h}{\partial \mu_{\dot\alpha}}\,\frac{\partial}{\partial\mu^{\dot\alpha}}\,, \qquad h\in\Omega^{0,1}(\PT,\cO(2))\,.
\ee
Integrability of the complex structure associated with $\nbar$ then imposes the equation
\be\label{intcond1}
\nbar^2=0 \quad \Leftrightarrow \quad \dbar h+\frac{1}{2}\left\{h,\,h\right\}=0\,,
\ee
where $\{\cdot,\,\cdot\}$ denotes the weighted Poisson bracket of \eqref{Pstruct}.

Rational curves in $\CPT$ which are holomorphic with respect to the deformed complex structure $\nbar$ have normal bundle $\cO(1)\otimes\C^{2m}$, and theorems of Kodaira~\cite{Kodaira:1962,Kodaira:1963} then imply that the moduli space of such curves is $4m$-dimensional. This moduli space of holomorphic twistor curves is the HK manifold associated to $\CPT$. 

To see this, let $x^{\alpha\dot\alpha}$ label the holomorphic curve $X\cong\P^1$, which is described by 
\be\label{holcurv}
\mu^{\dot\alpha}=F^{\dot\alpha}(x,\lambda)\,,
\ee
where $F^{\dot\alpha}(x,\lambda)$ is homogeneous of degree one in $\lambda_\alpha$ and obeys
\be\label{holcurveeq}
\dbar|_{X}F^{\dot\alpha}(x,\lambda)=\left.\frac{\partial h}{\partial\mu_{\dot\alpha}}\right|_{X}\,.
\ee
One can then construct the 2-form~\cite{Gindikin:1986}
\be\label{Gindikin2}
\d_x F^{\dot\alpha}\wedge\d_{x}F_{\dot\alpha}\,,
\ee
which is easily seen to be holomorphic on the twistor curve $X$. By an extension of Liouville's theorem, it then follows that 
\be\label{Gindikin3}
\d_x F^{\dot\alpha}\wedge\d_{x}F_{\dot\alpha}=\Sigma^{\alpha\beta}(x)\,\lambda_{\alpha}\,\lambda_{\beta}\,,
\ee
where $\Sigma^{\alpha\beta}$ is a triplet of 2-forms on the $4m$-dimensional moduli space. From the construction, it can be shown that $\Sigma^{\alpha\beta}=e^{\alpha\dot\alpha}\wedge e^{\beta}{}_{\dot\alpha}$ for some frame $e^{\alpha\dot\alpha}$, and that $\d\Sigma^{\alpha\beta}=0$, ensuring that the metric
\be\label{genKHmet}
\d s^{2}=\veps_{\alpha\beta}\,\veps_{\dot\alpha\dot\beta}\,e^{\alpha\dot\alpha}\,e^{\beta\dot\beta}\,,
\ee
is hyperk\"ahler. 

Note that \emph{real} structures can be obtained by imposing additional constraints on the twistor construction. For instance, Riemannian HK manifolds are obtained by equipping the twistor space with an anti-holomorphic involution fixing the twistor curves corresponding to Euclidean-real points in the manifold~\cite{Atiyah:1978wi,Hitchin:1986ea}, while split signature $(2m,2m)$ null-HK metrics can be obtained by taking a suitable real slice of $\CPT$~\cite{LeBrun:2005qf,Dunajski:2020qhh,Mason:2022hly}.


\subsection{Hyperholomorphic gauge fields \& their twistor theory}


Just as the hyperk\"ahler condition is the natural generalization of the self-dual vacuum Einstein equations to higher-dimensions, there is a natural generalization of the self-dual Yang-Mills equations. Consider a complexified, non-abelian gauge field on a HK manifold of dimension $4m$; the field strength is a 2-form valued in the Lie algebra of the gauge group. Exploiting the HK structure (in the notation used above), this 2-form can be decomposed into irreducibles as~\cite{Ward:1983zm}: 
\begin{equation}\label{Fdecomp}
F_{\alpha\dt{\alpha}\beta\dt{\beta}}=\veps_{\alpha\beta}\,\tilde{F}_{\dt{\alpha}\dt{\beta}}+\veps_{\dt{\alpha}\dt{\beta}}\,\hat{F}_{\alpha\beta}+\bv{F}_{\alpha\beta\,\dot\alpha\dot\beta}\,,
\end{equation}
where
\be\label{Freps1}
\tilde{F}_{\dot\alpha\dot\beta}=\frac{1}{2}\,\veps^{\alpha\beta}\,F_{\alpha\dot\alpha\beta\dot\beta}\,, \qquad \tilde{F}_{\dot\alpha\dot\beta}=\tilde{F}_{(\dot\alpha\dot\beta)}\,,
\ee
\be\label{Freps2}
\hat{F}_{\alpha\beta}=\frac{1}{2m}\,\veps^{\dot\alpha\dot\beta}\,F_{\alpha\dot\alpha\beta\dot\beta}\,, \qquad \hat{F}_{\alpha\beta}=\hat{F}_{(\alpha\beta)}\,,
\ee
and
\be\label{Freps3}
\bv{F}_{\alpha\beta\,\dot\alpha\dot\beta}=\bv{F}_{(\alpha\beta)[\dot\alpha\dot\beta]}\,, \qquad \veps^{\dt{\alpha}\dt{\beta}}\,\bv{F}_{\alpha\beta\,\dt{\alpha}\dt{\beta}}=0\,.
\ee
Note that when $m=1$, $\mathfrak{sp}(1,\C)\cong\mathfrak{sl}(2,\C)$ and there are no trace-free antisymmetric representations. In this case $\bv{F}_{\alpha\beta\,\dot\alpha\dot\beta}$ vanishes and $\tilde{F}_{\dot\alpha\dot\beta}$, $\hat{F}_{\alpha\beta}$ correspond to the self-dual and anti-self-dual parts of the field strength, respectively. However, for $m>1$ the trace-free irreducible $\bv{F}_{\alpha\beta\,\dot\alpha\dot\beta}$ is generically non-vanishing.

A gauge field is said to be \emph{hyperholomorphic} (HH) if~\cite{Corrigan:1982th,Ward:1983zm,Corrigan:1984si,Salamon:1984,Capria:1988,Verbitsky:1993} 
\be\label{HHcond}
\hat{F}_{\alpha\beta}=0=\bv{F}_{\alpha\beta\,\dot\alpha\dot\beta}\,,
\ee
meaning that its field strength is entirely encoded by the irreducible $\tilde{F}_{\dot\alpha\dot\beta}$ in the decomposition \eqref{Fdecomp}. This is equivalent to saying that the field strength is of type $(1,1)$ with respect to every complex structure of the HK manifold\footnote{This definition has been re-discovered multiple times through the years in both the mathematical physics and algebraic geometry literature, often being given different names by different authors. We use the modern mathematical nomenclature as it captures the close analogy with hyperk\"ahler geometry.}. When $m=1$, the HH conditions are equivalent to the self-dual Yang-Mills equations.

It is straightforward to show that any HH gauge field is automatically a solution of the Yang-Mills equations~\cite{Corrigan:1982th,Ward:1983zm,Corrigan:1984si}. This follows from the existence of a parallel 4-form on every HK manifold~\cite{Kraines:1966}, which induces an involution on the space of 2-forms under which each irreducible of the field strength has a different eigenvalue. For HH fields, this 4-form can be used to show that the Yang-Mills equations becomes equivalent to the Bianchi identity for the hyperholomorphic field strength.

\medskip

There is a twistor construction for HH gauge fields which generalizes Ward's original correspondence for self-dual gauge fields in 4-dimensions~\cite{Ward:1977ta}; this was first given by Ward for HH bundles on $\C^{4m}$~\cite{Ward:1983zm} and subsequently extended to generic HK manifolds~\cite{Kaledin:1998non}. The precise statement is:
\newline
\newline
\begin{thm}\label{Thm:WardC}
    There is a one-to-one correspondence between:
    \begin{itemize}
        \item hyperholomorphic bundles with gauge group $\GL(N,\C)$ on suitably convex regions of hyperk\"ahler $4m$-manifolds, and
        \item holomorphic vector bundles $E\to\CPT$ of rank $N$ on the associated twistor space that are topologically trivial when restricted to any holomorphic curve with normal bundle $\cO(1)\otimes\C^{2m}$.
    \end{itemize}
\end{thm}
Here, we have not stated the theorem for the most general gauge group, but it can be suitably refined to obtain other gauge groups (with or without real structures) by imposing further conditions on the holomorphic vector bundle $E\to\CPT$ (cf., \cite{Atiyah:1977pw,Ward:1990vs}). 

In practical terms, the holomorphic structure on $E$ can be encoded in a partial connection $\bar{D}:\Omega^{0}(\CPT, E)\to\Omega^{0,1}(\CPT,E)$ obeying $\bar{D}^2=0$. Locally, such a partial connection takes the form
\be\label{pconn1}
\bar{D}=\nbar+\cA\,, \qquad \cA\in\Omega^{0,1}(\CPT,\cO\otimes\mathrm{End}\,E)\,,
\ee
and the condition that $E|_{X}$ is topologically trivial implies that, generically, there exists a holomorphic trivialization for $E|_{X}$~\cite{Sparling:1990,Ivancovich:1990gj,Mason:2010yk,Bullimore:2011ni}. This means that there generically exists a holomorphic frame $H:E|_{X}\to\C^N$ such that
\be\label{Holframe}
\bar{D}|_{X} H=0\,.
\ee
The HH gauge field on the HK manifold is then obtained from
\be\label{Holframe2}
H^{-1}\,\lambda^{\alpha}\,V_{\alpha\dot\alpha}\,H=-\im\,\lambda^{\alpha}\,A_{\alpha\dot\alpha}(x)\,,
\ee
by an extension of Liouville's theorem, where $V_{\alpha\dot\alpha}$ is the dual of the frame $e^{\alpha\dot\alpha}$ for the HK metric and $A_{\alpha\dot\alpha}$ is valued in the Lie algebra of the gauge group. It follows that $A_{\alpha\dot\alpha}$ transforms like a gauge potential on the HK manifold under gauge transformations of the holomorphic frame of $E|_X$. The field strength is then seen to be hyperholomorphic as a consequence of the integrability of $\bar{D}$ on twistor space. 


\section{Chiral algebras from twistor space}
\label{sec 3}
The description of HK gravitational fields and HH gauge fields in twistor space gives a natural framework to study the symmetries of these sectors. By `symmetries of the HK/HH sector,' we mean those perturbations around a given, fixed HK/HH solution of the Einstein/Yang-Mills equations which preserve the HK/HH conditions. In twistor space, these are simply the perturbations of the holomorphic structures underpinning Theorems~\ref{Thm:NLGrav} and~\ref{Thm:WardC} .

This is particularly straightforward when considering the symmetries of the HK/HH sector around the flat background (which is trivially HK or HH). In this section, we show that the set of deformations of the HK and HH sectors in this case can be expanded in a set of modes which form infinite-dimensional chiral algebras under the natural bracket structures on twistor space.


\subsection{The HK sector and $\cL\mathfrak{ham}(\C^{2m})$}
\label{Sec:HKalg}

Consider the twistor space $\PT$ of flat $\C^{4m}$ with its natural HK structure. Theorem~\ref{Thm:NLGrav} indicates that symmetries of the HK sector around this flat background are given by infinitesimal deformations of the complex structure of $\PT$ which preserve its holomorphic fibration over $\P^1$ and the Poisson structure on the $\C^{2m}$ fibres. Just as when $m=1$, $\PT$ can be covered by two open subsets which exclude the north and south poles of the $\P^1$ base:
\be\label{opensets}
U=\{Z\in\PT|\lambda_0\neq 0\}\,, \qquad \tilde{U}=\{Z\in\PT|\lambda_1\neq 0\}\,.
\ee
A \emph{finite} complex deformation corresponds to a non-trivial patching between these two open sets, now with coordinates $Z^A=(\mu^{\dot\alpha},\lambda_\alpha)$ and $\tilde{Z}^A=(\tilde{\mu}^{\dot\alpha},\tilde{\lambda}_{\alpha})$, respectively.

The requirement that such a deformation preserves the holomorphic fibration over $\P^1$ means that $\tilde{\lambda}_{\alpha}=\lambda_{\alpha}$, while preserving the weighted holomorphic Poisson structure $I$ on the fibres means that the patching can be defined implicitly by a generating function of canonical transformations on $\C^{2m}$:
\be\label{genfunc}
G(\lambda_{\alpha},\mu^{\dot{1}},\ldots,\mu^{\dot{m}},\tilde{\mu}^{\dot{m+1}},\ldots,\tilde{\mu}^{\dot{2m}})\,,
\ee
with
\be\label{genfunc2}
\begin{split}
\tilde{\mu}^{\dot{\alpha}} &=\veps^{\dot\alpha\dot\beta}\,\frac{\partial G}{\partial\tilde{\mu}^{\dot\beta}}\,, \quad \mbox{for }\, \dot\alpha=1,\ldots,m\,, \\
\mu^{\dot\alpha}&=-\veps^{\dot\alpha\dot\beta}\,\frac{\partial G}{\partial\mu^{\dot\beta}}\,, \quad \mbox{for }\, \dot\alpha=m+1,\ldots,2m\,.
\end{split}
\ee
Clearly, for the patching to be well-defined projectively, the generating function $G$ must be homogeneous of weight two.

Infinitesimal deformations, corresponding to symmetries of the HK sector around flat space, therefore correspond to Hamiltonian (with respect to the symplectic form $\veps^{\dot\alpha\dot\beta}$) functions $g(Z)=\delta G$, with $g(Z)$ taking values in $\cO(2)$ and defined on the intersection $U\cap\tilde{U}$ of the open cover of $\PT$. This means that $g(Z)$ must be polynomial in the $\mu^{\dot\alpha}$ twistor coordinates on the fibres of $\pi:\PT\to\P^1$, but can be a generic Laurent series in the coordinates on the Riemann sphere base. In other words, an infinitesimal deformation can be expanded as
\be\label{HKmode1}
g(Z)=\sum_{r\in\Z}\,\sum_{\mbf{p}\in\N^{2m}_0}c[\mbf{p};r]\,g[\mbf{p};r]\,,
\ee
where $c[\mbf{p};r]\in\C$ are numerical coefficients, $\N_0$ is the set of natural numbers including zero and the modes are defined by
\be\label{HKmodes2}
g[\mbf{p};r]:=\frac{(\mu^{\dot{1}})^{p_1}\,(\mu^{\dot{2}})^{p_2}\cdots(\mu^{\dot{2m}})^{p_{2m}}}{\lambda_0^{P-2-r}\,\lambda_1^{r}}
\ee
for
\be\label{HKmodes3}
\mathbf{p}=(p_1,\ldots,p_{2m})\in\N_{0}^{2m}\,, \qquad P:=\sum_{i=1}^{2m}p_i\in\N_0\,.
\ee
It is easy to verify that each of the modes $g[\mbf{p};r]$ is homogeneous of degree two, as required.

These modes form a closed algebra under the weighted Poisson bracket of $\PT$:
\be\label{Poialg1}
\left\{g[\mbf{p};r],\,g[\mbf{q};s]\right\}=\sum_{i=1}^m\left(p_i\,q_{i+m}-p_{i+m}\,q_i\right) g[\mbf{p}+\mbf{q}-\mbf{1}_i-\mbf{1}_{i+m};r+s]\,,
\ee
where 
\be\label{ident}
(\mbf{1}_i)_{j}=\delta_{ij}\,.
\ee
To understand how to interpret this algebra, note that writing
\be\label{Hammodes1}
g[\mbf{p};r]=\frac{w[\mathbf{p}]}{\lambda_0^{P-2-r}\,\lambda_1^{r}}\,,
\ee
one finds that $w[\mathbf{p}]$ form a basis for the Poisson-diffeomorphisms of $\C^{2m}$ with the algebra
\be\label{Hammodes2}
\left\{w[\mbf{p}],\,w[\mbf{q}]\right\}=\sum_{i=1}^m\left(p_i\,q_{i+m}-p_{i+m}\,q_i\right) w[\mbf{p}+\mbf{q}-\mbf{1}_i-\mbf{1}_{i+m}]\,,
\ee
being $\mathfrak{ham}(\C^{2m})$, the Lie algebra of Hamiltonian (with respect to the weighted symplectic structure dual to $I$) vector fields on $\C^{2m}$. Inclusion of rational dependence on the holomorphic coordinates of $\P^1$ then amounts to taking the complexification of the loop algebra of $\mathfrak{ham}(\C^{2m})$. 

Indeed, working on the patch $U$ where $\lambda_{0}\neq0$, one can consider
\be\label{affinepatch1}
Z^{A}\sim\lambda_0^{-1}\,Z^A=(\zeta^{\dot\alpha},\,1,\lambda)\,,
\ee
where $\zeta^{\dot\alpha}\equiv\mu^{\dot\alpha}/\lambda_0$ and $\lambda\equiv\lambda_1/\lambda_0$ are affine (non-homogeneous) coordinates on the patch. In this patch,
\be\label{Loopalg1}
g[\mathbf{p};r]=\lambda_0^2\,\frac{(\zeta^{\dot{1}})^{p_1}\cdots(\zeta^{\dot{2m}})^{p_{2m}}}{\lambda^r}\equiv\frac{w[\mathbf{p}]}{\lambda^r}\,,
\ee
with $\lambda\in\C$ playing the role of a complexified loop parameter. We will often abuse notation by using the same symbols (e.g., $w[\mathbf{p}]$) to denote quantites in both homogeneous and affine coordinates.

Thus, the algebra of infinitesimal deformations of the HK structure of $4m$-dimensional flat space forms the algebra $\cL\mathfrak{ham}(\C^{2m})$. On the `spacetime' $\C^{4m}$, these infinitesimal deformations correspond to linearised HK gravitational perturbations: that is, gravitons whose linearised curvature tensor is contained in the totally symmetric representation of $\mathfrak{sp}(m,\C)$. In fact, this is an example of a \emph{chiral algebra}, a meromorphic vertex algebra defined on the Riemann sphere. To see this, note that the modes $g[\mbf{p};r]$ can be defined as the coefficients in a formal Laurent series 
\be\label{HKLaur1}
g[\mbf{p}](z):=\frac{w[\mbf{p}]}{\lambda-z}=\sum_{r\in\Z}g[\mbf{p};r]\,z^{r-1}\,,
\ee
where $z$ is an affine coordinate on the Riemann sphere. It then follows that\footnote{Here, it is understood that the Poisson bracket evaluated in the affine patch is the one naturally inherited from the weighted Poisson bracket in homogeneous coordinates. In particular, the un-weighted Poisson structure on $\C^{2m}$ is simply
\be\label{affinepatch2}
I_{\C^{2m}}:=\veps^{\dot\alpha\dot\beta}\,\frac{\partial}{\partial\zeta^{\dot\alpha}}\wedge\frac{\partial}{\partial\zeta^{\dot\beta}}\,,
\ee
in the notation of \eqref{affinepatch1}.}
\begin{multline}\label{HKvoa}
\left\{g[\mbf{p}](z),\,g[\mbf{q}](z')\right\}=\frac{1}{z-z'}\,\sum_{i=1}^m\left(p_i\,q_{i+m}-p_{i+m}\,q_i\right) g[\mbf{p}+\mbf{q}-\mbf{1}_i-\mbf{1}_{i+m}](z') \\+O\!\left((z-z')^0\right)\,,
\end{multline}
which is precisely the form of an operator product expansion in a chiral algebra. In other words, HK gravitational perturbations to flat space form the chiral algebra $\cL\mathfrak{ham}(\C^{2m})$ under the Poisson bracket on twistor space.

\medskip

Each of the modes $g[\mbf{p};r]$ of the algebra $\cL\mathfrak{ham}(\C^{2m})$ corresponds to a linear HK gravitational field on $\C^{4m}$. The linearised metric perturbations corresponding to a particular mode are constructed through the usual contour integral formulae~\cite{Penrose:1969ae,Hitchin:1980hp,Alexandrov:2008ds}:
\be\label{HKmetpert}
\begin{split}
h_{\alpha\dot\alpha\beta\dot\beta}(x)&=\frac{\iota_{\alpha}\,\iota_{\beta}}{2\pi\im}\oint\frac{\D\lambda}{\la\lambda\,\iota\ra^2}\,\left.\frac{\partial^{2}g[\mbf{p};r]}{\partial\mu^{\dot\alpha}\,\partial\mu^{\dot\beta}}\right|_{X} \\
&= \frac{\iota_{\alpha}\,\iota_{\beta}}{2\pi\im}\oint\d\lambda\,\left.\frac{\partial^{2}g[\mbf{p};r]}{\partial\mu^{\dot\alpha}\,\partial\mu^{\dot\beta}}\right|_{X}\,,
\end{split}
\ee
where $\D\lambda=\la\lambda\,\d\lambda\ra$ is the holomorphic measure on the twistor line $X$. In these expressions, the contour integral is taken around poles on the twistor line (or, in the second line, the complex plane parametrized by $\lambda$) and the constant spinor $\iota_{\alpha}$ corresponds to a gauge choice for the metric perturbation. Taking $\iota_{\alpha}=(0,1)$, on the patch where $\lambda_{\alpha}=(1,\lambda)$ it follows that $\la \iota\,\lambda\ra=1$ and the integral formula can be written as a contour integral on the complex $\lambda$-plane.

To see that these metric perturbations are themselves HK, one can compute their linearised Riemann tensor. It is easy to show that the only non-vanishing irreducible is
\be\label{HKRiem}
\begin{split}
\tilde{\psi}_{\dot\alpha\dot\beta\dot\gamma\dot\delta}(x)&=\frac{1}{2\pi\im}\oint\D\lambda\,\left.\frac{\partial^{4}g[\mbf{p};r]}{\partial\mu^{\dot\alpha}\cdots\partial\mu^{\dot\delta}}\right|_{X} \\
&= \frac{1}{2\pi\im}\oint\d\lambda\,\left.\frac{\partial^{4}g[\mbf{p};r]}{\partial\mu^{\dot\alpha}\cdots\partial\mu^{\dot\delta}}\right|_{X}\,.
\end{split}
\ee
As required, this is the totally symmetric $\mathfrak{sp}(m,\C)$ irreducible, and dependence on the gauge choice $\iota_{\alpha}$ has completely dropped out.
 

\subsection{The HH sector and $\cL\mathfrak{g}[\C^{2m}]$}
\label{Sec:HHalg}

Consider non-abelian gauge fields on $\C^{4m}$ equipped with its natural HK structure; Theorem~\ref{Thm:WardC} states that HH bundles are described by holomorphic vector bundles\footnote{Throughout, we will be agnostic as to the particular choice of gauge group, working with GL$(N,\C)$ in the first instance. As ever, specific gauge groups (including SL$(N,\C)$, SU$(N)$, SO$(N)$ or Sp$(N)$) can be obtained by endowing the bundle $E\to\PT$ and $\PT$ itself with additional structures (cf.,\cite{Atiyah:1977pw,Atiyah:1979iu,Ward:1981kj,Ward:1990vs}).} $E\to\PT$ which are trivial on each twistor line $X$ corresponding to $x\in\C^{4m}$. Therefore, symmetries of the HH sector around the trivial configuration of flat gauge fields correspond to infinitesimal deformations of the holomorphic structure on a holomorphically and topologically trivial vector bundle $E\to\PT$.

Now, the holomorphic structure of $E\to\PT$ is generally encoded in a patching function $\cF(Z)$ which takes values in $\mathrm{End}\,E$ and is homogeneous of weight zero. This must be analytic on $U\cap\tilde{U}$ and can generically be split on twistor lines as~\cite{Ward:1977ta,Atiyah:1977pw,Ward:1981kj}
\be\label{Wardsplit}
\cF(Z)|_{X}=\tilde{H}(x,\lambda)\,H^{-1}(x,\lambda)\,,
\ee
for matrices $H$, $\tilde{H}$ which are analytic on $U$ and $\tilde{U}$, respectively. These matrices give holomorphic frames for $E\to\PT$, in the sense that they obey \eqref{Holframe} and serve to define the associated partial connection $\bar{D}$ on $E$.

For a trivial gauge field, this partial connection is simply $\bar{D}=\dbar$ and the patching function is trivial: $\cF(Z)|_X=\mathrm{id}$. Therefore, infinitesimal deformations, corresponding to symmetries of the HH sector around flat gauge fields, correspond to functions $a(Z)$ homogeneous of weight zero, defined on the intersection $U\cap\tilde{U}$ and valued in endomorphisms of a (topologically and holomorphically) trivial bundle over $\PT$. This means that $a(Z)$ must be a polynomial in $\mu^{\dot\alpha}$, a generic Laurent series in $\lambda_{\alpha}$, and valued in $\mathfrak{g}$, the Lie algebra of the gauge group, since $\mathrm{End}\,E$ can be identified with $\mathfrak{g}$ for the trivial bundle.

This gives an infinitesimal deformation of the patching function on $E$ which can be expanded in modes
\be\label{HHmode1}
a^{\msf{a}}(Z)=\sum_{r\in\Z}\,\sum_{\mbf{p}\in\N_{0}^{2m}}c[\mbf{p};r]\,S^{\sa}[\mbf{p};r]\,,
\ee
where $\sa=1,\ldots,\dim\mathfrak{g}$ runs over the adjoint representation of the gauge group and the modes are defined by
\be\label{HHmode2}
S^{\sa}[\mbf{p};r]:=\msf{T}^{\sa}\,\frac{(\mu^{\dot{1}})^{p_1}\,(\mu^{\dot{2}})^{p_2}\cdots(\mu^{\dot{2m}})^{p_{2m}}}{\lambda_0^{P-r}\,\lambda_1^{r}}=\msf{T}^{\sa}\,\frac{(\zeta^{\dot{1}})^{p_1}\,(\zeta^{\dot{2}})^{p_2}\cdots(\zeta^{\dot{2m}})^{p_{2m}}}{\lambda^{r}}\,,
\ee
where $\msf{T}^{\sa}$ is a generator of the Lie algebra $\mathfrak{g}$. The second equality of \eqref{HHmode2} demonstrates that, as in the HK case, $\lambda$ can be viewed as a complexified loop parameter. These modes form an algebra under the commutator in the Lie algebra:
\be\label{Salg1}
\left[S^{\msf{a}}[\mbf{p};r],\,S^{\msf{b}}[\mbf{q};s]\right]=f^{\msf{abc}}\,S^{\msf{c}}[\mbf{p}+\mbf{q};r+s]\,,
\ee
for $f^{\msf{abc}}$ the structure constants of $\mathfrak{g}$.

Writing each mode as 
\be\label{HHmode3}
S^{\sa}[\mbf{p};r]=\frac{v^{\msf{a}}[\mbf{p}]}{\lambda_0^{P-r}\,\lambda_1^{r}}\,, \qquad v^{\msf{a}}[\mbf{p}]:=\msf{T}^{\sa}\,(\mu^{\dot{1}})^{p_1}\,(\mu^{\dot{2}})^{p_2}\cdots(\mu^{\dot{2m}})^{p_{2m}}\,,
\ee
the objects $v^{\msf{a}}[\mbf{p}]$ can be viewed as degree $P$ polynomial maps from the $\C^{2m}$ fibres of $\PT$ into $\mathfrak{g}$. Thus, the algebra \eqref{Salg1} is the loop algebra of the Lie algebra of polynomial maps $\C^{2m}\to\mathfrak{g}$, denoted $\cL\mathfrak{g}[\C^{2m}]$. This is again a chiral algebra, as is apparent by defining the formal Laurent series
\be\label{HHLaur1}
S^{\sa}[\mbf{p}](z):=\frac{v^{\sa}[\mbf{p}]}{\lambda-z}=\sum_{r\in\Z}S^{\sa}[\mbf{p};r]\,z^{r-1}\,.
\ee
for $z$ an affine coordinate on the Riemann sphere. It is then straightforward to show that
\be\label{HHvoa}
\left[S^{\sa}[\mbf{p}](z),\,S^{\msf{b}}[\mbf{q}](z')\right]=\frac{f^{\msf{abc}}}{z-z'}\,S^{\msf{c}}[\mbf{p+q}](z')+O\!\left((z-z')^0\right)\,,
\ee
as expected for a chiral vertex operator algebra.

\medskip

Each mode $S^{\sa}[\mbf{p};r]$ of $\cL\mathfrak{g}[\C^{2m}]$ corresponds to a linear HH gauge field, or gluon, on $\C^{4m}$. As in the gravitational case, it is straightforward to recover the linear gauge field from each mode via a contour integral formula~\cite{Penrose:1969ae,Sparling:1990}:
\be\label{HHgfpert}
\begin{split}
a^{\sa}_{\alpha\dot\alpha}(x)&=\frac{\iota_{\alpha}}{2\pi\im}\oint\frac{\D\lambda}{\la\lambda\,\iota\ra}\,\left.\frac{\partial S^{\sa}[\mbf{p};r]}{\partial\mu^{\dot\alpha}}\right|_{X} \\
&= \frac{\iota_{\alpha}}{2\pi\im}\oint\d\lambda\,\left.\frac{\partial S^{\sa}[\mbf{p};r]}{\partial\mu^{\dot\alpha}}\right|_{X}\,,
\end{split}
\ee
with the choice of constant spinor $\iota_{\alpha}=(0,1)$ again corresponding to a choice of gauge for the HH gluon. It is easy to show that the linearised field strength associated to \eqref{HHgfpert} has only the HH irreducible under the decomposition \eqref{Fdecomp}, corresponding to
\be\label{HHfspert}
\begin{split}
\tilde{f}_{\dot\alpha\dot\beta}(x)&=\frac{1}{2\pi\im}\oint\D\lambda\,\left.\frac{\partial^{2} S^{\sa}[\mbf{p};r]}{\partial\mu^{\dot\alpha}\partial\mu^{\dot\beta}}\right|_{X} \\
 &= \frac{1}{2\pi\im}\oint\d\lambda\,\left.\frac{\partial^{2} S^{\sa}[\mbf{p};r]}{\partial\mu^{\dot\alpha}\partial\mu^{\dot\beta}}\right|_{X}\,,
\end{split}
\ee
as required.


\section{Twistor sigma models and chiral algebras}
\label{sec 4}

We have seen how the chiral algebras $\mcal{L}\mfk{ham}(\mbb{C}^{2m})$ and $\mcal{L}\mfk{g}[\mbb{C}^{2m}]$ emerge as the symmetry algebras of the hyperk\"ahler (HK) and hyperholomorphic (HH) sectors of gravity and gauge theory around flat space in $4m$-dimensions. These algebras are formed by the natural structures on twistor space describing the HK and HH sectors: a weighted holomorphic Poisson structure in the HK case, and the Lie bracket associated to the gauge algebra in the HH case. In this sense, the algebras appear almost kinematically from twistor space.

However, these chiral algebras can also be seen to emerge directly from dynamical descriptions of the HK and HH sectors in terms of certain classical 2d CFTs, referred to as \emph{twistor sigma models}. These give variational principles for the underlying holomorphic objects in the twistor construction. In the HK case, these are the holomorphic curves in twistor space corresponding to points in the HK manifold, while in the HH setting they are holomorphic frames encoding the HH gauge field on spacetime. By considering deformations of these holomorphic structures, vertex operators corresponding to HK/HH perturbations are obtained in these twistor sigma models. 

The chiral symmetry algebras can then be realized in the twistor sigma models by defining charges associated to the algebra modes. These charges coincide with the vertex operators associated to each mode via the \v Cech-Dolbeault correspondence, and their semi-classical algebra under the operator product expansion (OPE) of the twistor sigma model forms the chiral symmetry algebra.


\subsection{Gravity}

The HK metric associated to a twistor space $\CPT$ is encoded in its holomorphic rational curves, though the construction \eqref{holcurv} -- \eqref{genKHmet}. In terms of the Hamiltonian $h(Z)$ defining the deformed complex structure on $\CPT$, these curves are determined by the differential equation \eqref{holcurveeq}:
\be\label{holcurveeq2}
\dbar|_{X}F^{\dot\alpha}(x,\lambda)=\left.\frac{\partial h}{\partial\mu_{\dot\alpha}}\right|_{X}\,.
\ee
The twistor sigma model for the corresponding $4m$-dimensional HK structure is an action whose Euler-Lagrange equations are \eqref{holcurveeq2}, giving these holomorphic curves. This takes the form of a chiral, 2d classical CFT~\cite{Adamo:2021bej}:
\be\label{HKTS}
S[\mu^{\dot\alpha}]=\frac{1}{\hbar}\int_{X}\frac{\D\lambda}{\la\lambda\,o\ra^2\,\la\lambda\,\iota\ra^2}\left(\mu^{\dot\alpha}\,\dbar|_{X}\mu_{\dot\alpha}-2\,h|_{X}\right)\,,
\ee
where $\hbar$ is a formal parameter; the integral is over the Riemann sphere $X\cong\P^1$; $\mu^{\dot\alpha}=F^{\dot\alpha}(x,\lambda)$ are sections of $\cO(1)\to\P^1$; and $h|_{X}=h(\mu=F,\lambda)$. It is easy to see that \eqref{holcurveeq2} are the equations of motion for this action. The constant spinors $o^{\alpha}$, $\iota^{\alpha}$ -- normalized so that $\la\iota\,o\ra=1$ -- appearing in the measure of \eqref{HKTS} amount to a gauge choice for the resulting holomorphic curves, namely that $F^{\dot\alpha}(x,\lambda)$ has first-order zeros at $\lambda^{\alpha}=o^{\alpha},\iota^{\alpha}$. For consistency, this also requires that the Hamiltonian $h|_{X}$ have second order zeros at these points on the Riemann sphere\footnote{It is possible to reformulate the twistor sigma model in a way that does not explicitly break M\"obius invariance in this way, by working with curves of homogeneity $-1$~\cite{Adamo:2021bej}. Here, we prefer to work with positive degree to make closer contact with constructions more familiar in the scattering amplitudes literature.}.

Beyond providing a variational principle for the holomorphic curves in $\CPT$, the twistor sigma model directly encodes the HK structure on spacetime. Indeed, it can be shown that evaluated on-shell -- that is, on solutions of \eqref{holcurveeq2} -- the twistor sigma model is related to a potential $\Omega(x)$ for the HK metric~\cite{Adamo:2021bej}:
\be\label{Plebpot1}
\Omega(x)=x^2-\frac{\hbar}{4\pi\im}\,S[F]\,,
\ee
where $S[F]$ indicates the action \eqref{HKTS} evaluated on $\mu^{\dot\alpha}=F^{\dot\alpha}(x,\lambda)$ a solution of \eqref{holcurveeq2}. The scalar $\Omega(x)$ is known as the first Plebanski potential~\cite{Plebanski:1975wn}, and defines a frame
\be\label{HKframe}
e^{\alpha\dot\alpha}=\left(o_{\alpha}\,\d x^{\alpha\dot\alpha},\,-o^{\gamma}\,\iota^{\delta}\,\frac{\partial^{2}\Omega}{\partial x^{\gamma}{}_{\dot\alpha}\partial x^{\delta\dot\beta}}\,\iota_{\alpha}\,\d x^{\alpha\dot\beta}\right)\,.
\ee
That the metric \eqref{genKHmet} associated to this frame is HK follows from the fact that $\Omega(x)$ obeys
\be\label{1Plebeq}
\det\!\left(o^{\alpha}\,\iota^{\beta}\,\frac{\partial^{2}\Omega}{\partial x^{\alpha\dot\alpha}\partial x^{\beta\dot\beta}}\right)=2\,,
\ee
a condition often referred to as `the first heavenly equation'~\cite{Plebanski:1975wn}.

\medskip

Thus, the twistor sigma model \eqref{HKTS} directly encodes the HK structure on spacetime, both through its equations of motion and its on-shell values. As we saw in Section~\ref{Sec:HKalg}, an infinitesimal deformation of the HK structure corresponds on twistor space to a Hamiltonian $g(Z)$ which is homogeneous of weight two and polynomial in $\mu^{\dot\alpha}$ but Laurent in $\lambda_{\alpha}$. This acts as
\be\label{algactws}
\delta \mu^{\dot\alpha}=\left\{g,\,\mu^{\dot\alpha}\right\}=\veps^{\dot\beta\dot\alpha}\,\frac{\partial g}{\partial\mu^{\dot\beta}}\,, \qquad \delta h=\nbar g=\dbar g+\left\{h,\,g\right\}\,,
\ee
so we see that $g$ is only a symmetry of the twistor sigma model -- and thus preserves the HK metric on spacetime -- if it is holomorphic with respect to the complex structure on $\CPT$: $\delta h=\nbar g=0$.

For $\C^{4m}$, where $h=0$, this means that $g$ only preserves the flat HK metric if $\dbar g=0$; that is, if $g$ is holomorphic. However, a generic infinitesimal deformation is Laurent, rather than holomorphic, in the $\P^1$ base of $\PT$, as the mode expansion \eqref{HKmode1} indicates. Therefore, a generic deformation does \emph{not} preserve the metric: it is not a true symmetry of the twistor sigma model and thus alters the potential $\Omega$ and hence the HK metric (although the deformation is still HK). This is, of course, what we expect: $\cL\mathfrak{ham}(\C^{2m})$ are symmetries of the HK sector around flat space, not of the HK metric of $\C^{4m}$ itself. For instance,
\be\label{Translation}
g_{b}(Z)=b^{\alpha}{}_{\dot\alpha}\,\lambda_{\alpha}\,\mu^{\dot\alpha}\,,
\ee
obeys $\dbar g_b=0$ and is a symmetry of the twistor sigma model, but corresponds on spacetime simply to the translation generated by the vector $b^{\alpha\dot\alpha} \partial_{\alpha\dot\alpha}$.

So, given a generic Hamiltonian deformation $g$ of $\PT$, one can define an associated charge
\be\label{gCharge1}
Q_g=\frac{1}{2\pi\im}\oint g\,\frac{\D\lambda}{\la\lambda\,o\ra^2\,\la\lambda\,\iota\ra^2}\,,
\ee
with $g$ understood to be pulled back to $\P^1$ via the incidence relation \eqref{incidence} and the contour taken around poles\footnote{Observe that the poles from $\la\lambda\, o\ra=0=\la\lambda\,\iota\ra$ in the measure do not contribute, since by assumption all data (including the Hamiltonian $g$) must have second order zeros at these points for the twistor sigma model to be well-defined.}. This charge will be conserved only if $g$ is globally holomorphic (in which case it is simply a symmetry of $\C^{4m}$).

The algebra of these charges is determined by the free, classical theory
\be\label{TSMflat}
S^{\mathrm{free}}[\mu]=\int_{X}\frac{\D\lambda}{\la\lambda\,o\ra^2\,\la\lambda\,\iota\ra^2}\wedge \mu^{\dot\alpha}\,\dbar|_X\mu_{\dot\alpha}\,,
\ee
which has the simple OPE
\be\label{muOPE}
\mu^{\dot\alpha}(\lambda)\,\mu^{\dot\beta}(\lambda')\sim\frac{\veps^{\dot\alpha\dot\beta}}{\la\lambda\,\lambda'\ra}\,\la o\,\lambda\ra\,\la\iota\,\lambda\ra\,\la o\,\lambda'\ra\,\la\iota\,\lambda'\ra\,,
\ee
where dependence on $x^{\alpha\dot\alpha}$ has been suppressed. With this, one finds that the semi-classical OPE between charges corresponds to the canonical commutation relations:
\begin{multline}\label{Qgcomm}
\left[Q_g,\,Q_g'\right]=\frac{1}{(2\pi\im)^2}\oint\oint\frac{\{g,\,g'\}}{\la\lambda\,\lambda'\ra}\,\frac{\D\lambda\,\D\lambda'}{\la o\,\lambda\ra\,\la\iota\,\lambda\ra\,\la o\,\lambda'\ra\,\la\iota\,\lambda'\ra} \\
=\frac{1}{2\pi\im}\oint \{g,\,g'\}\,\frac{\D\lambda}{\la o\,\lambda\ra^2\,\la\iota\,\lambda\ra^2}=Q_{\{g,g'\}}\,,
\end{multline}
with the semi-classical constraint meaning that only a single Wick contraction between the charges is allowed.

It then follows immediately that if we define the charge
\be\label{Qgmode1}
Q[\mbf{p};r]=\frac{1}{2\pi\im}\oint g[\mbf{p};r]\,\frac{\D\lambda}{\la\lambda\,o\ra^2\,\la\lambda\,\iota\ra^2}\,,
\ee
associated to a given mode, the semi-classical OPE between these charges gives
\be\label{Qgmode2}
\begin{split}
\left[Q[\mathbf{p};r],\,Q[\mbf{q};s]\right]&=Q_{\{g[\mbf{p};r],g[\mbf{q};s]\}} \\
&=\sum_{i=1}^{m}\left(p_i\,q_{i+m}-p_{i+m}\,q_i\right) Q[\mbf{p}+\mbf{q}-\mbf{1}_i-\mbf{1}_{i+m};r+s]\,,
\end{split}
\ee
which is precisely the chiral algebra $\cL\mathfrak{ham}(\C^{2m})$. Thus, as expected, the chiral symmetry algebra of the HK sector around flat space is captured by the algebra of associated charges in the twistor sigma model under OPE.

Note that there is a dictionary between the charges $Q_g$ and vertex operators in the twistor sigma model representing the HK gravitational perturbations. These vertex operators are given by
\be\label{gravVOs}
V_h=\int \frac{\D\lambda}{\la\lambda\,o\ra^2\,\la\lambda\,\iota\ra^2}\wedge h|_X\,, \qquad h\in H^{0,1}(\PT,\cO(2))\,,
\ee
where $h$ now represents an infinitesimal perturbation to the complex structure on $\PT$. Any such $h$ has a representative in \v{C}ech cohomology (by the \v{C}ech-Dolbeault isomorphism). For instance, on $U$, $h=\dbar g_0$ by the Poincar\'e lemma, and similarly $h=\dbar g_1$ on $\tilde{U}$, for $g_0$ and $g_1$ smooth functions of homogeneity two. Then $g=g_0-g_1$ is holomorphic on $U\cap\tilde{U}$ and defined up to the addition of holomorphic functions extending across both open sets; this means that $g$ defines a cohomology class in the \v{C}ech cohomology group $\check{H}^{1}(\PT,\cO(2))$ and can be expanded in the modes $g[\mbf{p};r]$. This \v{C}ech-Dolbeault isomorphism gives an equivalence between the vertex operators \eqref{gravVOs} and charges \eqref{gCharge1} $Q_g=V_h$ which holds inside the contour of integration for the charge.


\subsection{Gauge theory}

The HH sector of gauge theory on $\C^{4m}$ is described by a twistor sigma model which is classically equivalent to a worldsheet current algebra pulled back to twistor lines~\cite{Nair:1988bq,Witten:2003nn,Berkovits:2004hg}:
\be\label{HHtsm1}
S[\rho,\bar{\rho}]=\int_{X}\bar{\rho}_{i}\,\bar{D}|_{X}\rho^{i}\,,
\ee
where $\bar{D}|_{X}$ is the partial connection \eqref{pconn1} on the bundle $E\to\PT$ pulled back to a twistor line $X\cong\P^1$; and $\rho^{i}$, $\bar{\rho}_i$ are fermionic sections of $K^{1/2}_{\P^1}$ valued in the fundamental and anti-fundamental representations of the gauge group, respectively. Extrema of this action are simply those fermions which are holomorphic with respect to the partial connection, pulled back to $\P^1$. 

To see what this has to do with HH gauge fields, write 
\be\label{fermdecomp1}
\bar{\rho}_i(\lambda)=\frac{\sqrt{\D\lambda}}{\la\lambda\,o\ra}\,\alpha_i(\lambda)\,, \qquad \rho^i(\lambda)=\frac{\sqrt{\D\lambda}}{\la\lambda\,\iota\ra}\,\beta^{i}(\lambda)\,,
\ee
where the fermionic scalars $\alpha_i$ and $\beta^i$ have simple zeros at the points $\lambda_{\alpha}=o_{\alpha}$ and $\lambda_{\alpha}=\iota_{\alpha}$, respectively. One can then construct a matrix
\be\label{frameTS1}
H^{i}{}_{j}(x,\lambda)=\beta^{i}\,\alpha_j\,,
\ee
which automatically obeys $\bar{D}|_{X} H=0$, the equation of motion \eqref{Holframe} for a holomorphic frame of $E\to\PT$ from which the HH gauge field is constructed via \eqref{Holframe2}.

As we saw in Section~\ref{Sec:HHalg}, infinitesimal deformations of a HH gauge field are given on twistor space by functions $a(Z)$ valued in endomorphisms of $E\to\PT$ which are homogeneous of weight zero, polynomial in $\mu^{\dot\alpha}$ and Laurent in $\lambda_{\alpha}$. This effectively defines an infinitesimal gauge transformation
\be\label{tsgtrans}
\bar{D}\to\bar{D}+\bar{D}a\,,
\ee
from which we observe that $a$ is only a symmetry of \eqref{HHtsm1} if it is holomorphic with respect to the partial connection on $E$: $\bar{D}a=0$.

Around a trivial (flat) HH field configuration, where $\mathrm{End}\,E\cong\mathfrak{g}$ trivially, this means that $a^{\msf{a}}(Z)=\msf{T}^{\sa}\,a(Z)$ is only a true symmetry of the twistor sigma model if $\dbar a^{\sa}=0$, or $a^\sa(Z)$ is holomorphic. Thus, just as we saw in the gravitational case, a generic deformation of the trivial HH structure is not a symmetry of the twistor sigma model. 

Now, from the fields $\bar{\rho}_i$, $\rho^i$ of the twistor sigma model, we can form the Kac-Moody current
\be\label{KMcurr}
j^{\sa}(\lambda):=(\msf{T}^{\sa})^{i}{}_{j}\,\bar{\rho}_i(\lambda)\,\rho^{j}(\lambda)\,,
\ee
where $\msf{T}^{\sa}$ is a generator of $\mathfrak{g}$ and normal ordering for the fermions is implicit. Using this, the charge associated to any HH perturbation $a(Z)$ can be defined as
\be\label{aCharge1}
Q_{a}=\frac{1}{2\pi\im}\oint j^{\sa}\,a\,,
\ee
with $a$ pulled back to $\P^1$ via the incidence relations \eqref{incidence} and the contour taken around poles. This is a conserved charge only if $a(Z)$ is holomorphic.

The algebra of these charges around a flat gauge field configuration on $\C^{4m}$ is now determined by the free classical theory
\be\label{HHTSMfree}
S^{\mathrm{free}}[\rho,\bar{\rho}]=\int_{X}\bar{\rho}_{i}\,\dbar|_{X}\rho^{i}\,,
\ee
for which the fermions have the OPE
\be\label{rhoOPE}
\bar{\rho}_i(\lambda)\,\rho^{j}(\lambda')\sim \delta^{j}_{i}\,\frac{\sqrt{\D\lambda\,\D\lambda'}}{\la\lambda\,\lambda'\ra}\,.
\ee
This in turn induces the \emph{semi-classical} OPE 
\be\label{jjOPE}
j^{\sa}(\lambda)\,j^{\msf{b}}(\lambda')\sim f^{\msf{abc}}\,\frac{j^{\msf{c}}(\lambda')}{\la\lambda\,\lambda'\ra}\,\frac{\la\iota\,\lambda'\ra}{\la\iota\lambda\ra}\,\D\lambda\,,
\ee
where the rational factor depending on $\iota_{\alpha}=(0,1)$ is a gauge choice corresponding to inverting the $\dbar|_{X}$-operator. Note the absence of a double pole in \eqref{jjOPE}, which would typically be associated with the level of the Kac-Moody algebra. This is because we only consider the semi-classical OPE; the double pole arises from a double Wick contraction between the fermionic constituents of the Kac-Moody currents and is quantum mechanical in nature. Such multiple Wick contractions introduce multi-trace terms into current correlators of the Kac-Moody algebra which are indicative of gravitational couplings, while we are interested in the pure gauge theory. More generally, these contributions typically produce non-unitary, higher-derivative gravitational degrees of freedom (cf., \cite{Berkovits:2004jj,Azevedo:2017lkz,Azevedo:2018dgo,Adamo:2018hzd}). 

Armed with these OPEs, it is straightforward to compute the algebra of charges
\begin{multline}\label{Qacomm}
\left[Q_{a},\,Q_{a'}\right]=\frac{1}{(2\pi\im)^2}\oint\oint f^{\msf{abc}}\,j^{\msf{c}}(\lambda')\,\frac{\D\lambda\,\la\iota\,\lambda'\ra}{\la\lambda\,\lambda'\ra\,\la\iota\,\lambda\ra}\,a(\lambda)\,a'(\lambda') \\
=\frac{f^{\msf{abc}}}{2\pi\im}\oint a\,a'\,j^{\msf{c}}=Q_{[a,a']}\,.
\end{multline}
It then follows that for the charges associated to the modes of any $a(z)$
\be\label{Qamode1}
Q^{\sa}[\mbf{p};r]=\frac{1}{2\pi\im}\oint (S^{\sa}[\mbf{p};r])^{i}{}_{j}\,\bar{\rho}_i\,\rho^{j}\,,
\ee
one obtains 
\be\label{Qamode2}
\left[Q^{\sa}[\mbf{p};r],\,Q^{\msf{b}}[\mbf{q};s]\right]=f^{\msf{abc}}\,Q^{\msf{c}}[\mbf{p}+\mbf{q};r+s]\,,
\ee
which is the chiral algebra $\cL\mathfrak{g}[\C^{2m}]$, as expected.

In addition, there is a dictionary between the charges $Q_{a}$ and vertex operators in the twistor sigma model corresponding to HH gluons. This mirrors the \v{C}ech-Dolbeault isomorphism already discussed in the gravitational setting, with the vertex operators given by
\be\label{gluVO}
V_\cA=\int j^{\sa}\wedge\cA\,, \qquad \cA\in H^{0,1}(\PT,\cO)\,,
\ee
with $\cA$ being the Dolbeault representative of the $a(Z)$ in \v{C}ech cohomology.


\section{Conformally soft modes and holomorphic collinear limits}
\label{sec 5}

In 4d, the twistor chiral algebras $\cL\mathfrak{ham}(\C^2)$ and $\cL\mathfrak{g}[\C^2]$ were first discovered by considering gravitons and gluons in a \emph{conformal primary basis}~\cite{Pasterski:2016qvg,Pasterski:2017kqt}, where wavefunctions transform like conformal primaries on the celestial 2-sphere of null directions. It was then shown that the conformally soft tower of positive helicity gravitons or gluons~\cite{Donnay:2018neh,Banerjee:2019aoy,Fan:2019emx,Pate:2019mfs,Nandan:2019jas,Adamo:2019ipt,Guevara:2019ypd,Puhm:2019zbl} form a chiral algebra under holomorphic collinear limits~\cite{Guevara:2021abz}; equivalence with $\cL\mathfrak{ham}(\C^2)$ and $\cL\mathfrak{g}[\C^2]$ emerges in this framework only after a rather non-trivial re-labeling of the conformally soft modes~\cite{Strominger:2021mtt}. Chiral algebras are natural in the context of celestial holography, with the holomorphic collinear limit coinciding with the holomorphic OPE limit on the celestial 2-sphere~\cite{Fan:2019emx,Pate:2019lpp}.

However, in $d>4$ dimensions, where the celestial sphere is $S^{d-2}$, there is no clear reason why a chiral algebra should emerge from some notion of collinear limits. Indeed, for general $d$, $S^{d-2}$ does not even admit an almost complex structure. In this section, we show that for $d=4m$, the complex massless momenta of HK or HH modes are parametrized by a $\P^1\times\P^{2m-1}\subset S^{4m-2}_{\C}$, inside the complexified celestial sphere $S^{4m-2}_{\C}$. This gives rise to a notion of `holomorphic' collinear limit which is intrinsically one complex-dimensional, providing a direct link between the twistorial chiral algebras $\cL\mathfrak{ham}(\C^{2m})$/$\cL\mathfrak{g}[\C^{2m}]$ and algebras of conformally soft gravitons/gluons under this holomorphic collinear limit.


\subsection{HK/HH kinematics}

Making use of the underlying HK structure, generic gluon or graviton momentum eigenstates in $4m$-dimensional flat space can be written as
\be\label{glu/gravsf1}
a^{\sa}_{\alpha\dot\alpha}=\msf{T}^{\sa}\,\epsilon_{\alpha\dot\alpha}\,\e^{\im\,k\cdot x}\,, \qquad h_{\alpha\dot\alpha\beta\dot\beta}=\epsilon_{\alpha\dot\alpha}\,\epsilon_{\beta\dot\beta}\,\e^{\im\,k\cdot x}\,,
\ee
where $k^2=0=k\cdot\epsilon$ and $\epsilon\cdot\epsilon=0$. Since the HK structure of $\C^{4m}$ is complexified from Euclidean signature, we assume that the momenta are also complex (as there are no Euclidean-real null vectors). 

Consider the momentum $k^{\alpha\dot\alpha}$ itself. Just as in 4-dimensions, it is clear that any $4m$-vector which is \emph{simple}
\be\label{snv}
k_{\mathrm{simple}}^{\alpha\dot\alpha}=\kappa^{\alpha}\,\tilde{\kappa}^{\dot\alpha}\,,
\ee
will be null, since
\be\label{nullvec1}
k_{\mathrm{simple}}^{2}=\veps_{\alpha\beta}\,\veps_{\dot\alpha\dot\beta}\,\kappa^{\alpha}\,\tilde{\kappa}^{\dot\alpha}\,\kappa^{\beta}\,\tilde{\kappa}^{\dot\beta}=\la\kappa\,\kappa\ra\,[\tilde{\kappa}\,\tilde{\kappa}]=0\,,
\ee
by the skew-symmetry of $\veps_{\alpha\beta}$ and $\veps_{\dot\alpha\dot\beta}$. However, unlike in 4-dimensions, not every null vector in $4m$-dimensions takes this form for $m>1$. This is easy to see on dimensional grounds: the space of complexified null momenta in $4m$-dimensions is $(4m-1)$-dimensional, while the space of simple null momenta is $(2m+1)$-dimensional\footnote{This counting is given by the $2$ components of $\kappa^{\alpha}$ plus the $2m$ components of $\tilde{\kappa}^{\dot\alpha}$, minus the overall projective rescaling $\kappa^\alpha\to r\,\kappa^\alpha$, $\tilde{\kappa}^{\dot\alpha}\to r^{-1}\,\tilde{\kappa}^{\dot\alpha}$ which leaves $k^{\alpha\dot\alpha}$ invariant.}. In particular, for $m>1$, there are vectors $k^{\alpha\dot\alpha}$ which are null with respect to the symplectic form $\veps_{\dot\alpha\dot\beta}$ of Sp$(m,\C)$ but not $\veps_{\alpha\beta}$.

A general null vector can then be parametrized as
\be\label{gennv1}
k^{\alpha\dot\alpha}=\kappa^{\alpha\,a}\,\tilde{\kappa}^{\dot\alpha}_{a}\,, \qquad a=1,2\,,
\ee
subject to the constraint $[\tilde{\kappa}_1\,\tilde{\kappa}_2]=0$. Note that when $m>1$, this does \emph{not} force $\tilde{\kappa}_1^{\dot\alpha}$ to be proportional to $\tilde{\kappa}_{2}^{\dot\alpha}$. Now, there are initially $4m+4$ degrees of freedom in the parametrization \eqref{gennv1} ($4$ from $\kappa^{\alpha\,a}$ and $4m$ from $\tilde{\kappa}^{\dot\alpha}_{a}$), from which we must subtract 4 due to the invariance under GL$(2,\C)$ transformations
\be\label{gennv2}
\kappa^{\alpha\,a}\to\kappa^{\alpha\,b}\,\Lambda^{a}{}_{b}\,, \qquad \tilde{\kappa}_{a}^{\dot\alpha}\to(\Lambda^{-1})_{a}{}^{b}\,\tilde{\kappa}^{\dot\alpha}_{b}\,,
\ee
as well as another one degree of freedom due to the constraint $[\tilde{\kappa}_1\,\tilde{\kappa}_2]=0$. This leaves $4m-1$ degrees of freedom in \eqref{gennv1}, precisely the number required for a general null vector in $4m$-dimensions.

Next, let us consider the structure of a polarization vector $\epsilon_{\alpha\dot\alpha}$ associated to a HH gluon. In doing this, it makes sense to demand that the resulting polarization should be consistent with the case of a simple momentum \eqref{snv}, where the structure of the polarization is inherited from that of a positive helicity gluon in 4-dimensions (cf., \cite{DeCausmaecker:1981jtq,Kleiss:1985yh,Xu:1986xb,Berends:1987cv,Chalmers:1997ui,Witten:2003nn}):
\be\label{snvpol1}
\epsilon_{\alpha\dot\alpha}^{\mathrm{HH}}\Big|_{\mathrm{simple}}=\frac{\iota_{\alpha}\,\tilde{\kappa}_{\dot\alpha}}{\la\kappa\,\iota\ra}\,,
\ee
where the choice of constant SL$(2,\C)$ spinor $\iota_{\alpha}$ simply corresponds to fixing the residual gauge freedom in terms of a lightfront gauge condition. 

Consistency with \eqref{snvpol1} in the case of simple momenta then means that a HH polarization vector should take the form
\be\label{HHpol1}
\veps_{\alpha\dot\alpha}^{\mathrm{HH}}=\epsilon^{a}_{\alpha}\,\tilde{\kappa}_{\dot\alpha\,a}\,,
\ee
for some $\epsilon^{a}_{\alpha}$. Now, the HH conditions \eqref{HHcond} on the linearized field strength associated to such a polarization lead to the conditions
\be\label{HHpol2}
\kappa^{a}_{(\alpha}\,\epsilon^{b}_{\beta)}\,[\tilde{\kappa}_{a}\,\tilde{\kappa}_{b}]=0\,, \qquad \kappa^{a}_{(\alpha}\,\epsilon^{b}_{\beta)}\,\tilde{\kappa}_{[\dot\alpha\,|a|}\,\tilde{\kappa}_{\dot\beta]\,b}=0\,.
\ee
The first of these is trivially satisfied thanks to the null momentum constraint $[\tilde{\kappa}_1\,\tilde{\kappa}_2]=0$, but the second requires
\be\label{HHpol3}
\tilde{\kappa}_{[\dot\alpha\,|a|}\,\tilde{\kappa}_{\dot\beta]\,b}=0\,,
\ee
which sets $\tilde{\kappa}_{1}^{\dot\alpha}\propto \tilde{\kappa}_{2}^{\dot\alpha}$. In other words, the HH condition \emph{forces} the gluon momentum to be simple, of the form \eqref{snv}! It is straightforward to see that the same thing happens for HK gravitons.

At first, this might seem alarming, as the HK/HH condition restricts the kinematics of a graviton/gluon at the level of its momentum. However, a bit of thought reveals that this is actually to be expected. For $m>1$ the HH equations \eqref{HHcond} \emph{overconstrain} the underlying gauge field~\cite{Corrigan:1982th,Ward:1983zm}: this is because there are $3m(2m-1)$ equations while the gauge field itself has $4m$ components (before gauge-fixing). When $m>1$, this number of equations is always greater than the degrees of freedom in the gauge field, so the HH equations are an overconstrained system\footnote{This does \emph{not} mean that there are no interesting HH field configurations. Indeed, Theorem~\ref{Thm:WardC} shows that HH fields can be constructed from generic holomorphic data on twistor space, and there are also explicit ADHM-like constructions~\cite{Corrigan:1984si}.}. In other words, for $m>1$ the HH conditions do more than select one of the $4m-2$ on-shell polarizations of a gluon; they also constrain its momentum to lie on a codimension $2m-2$ subvariety of the null cone. The same story holds for HK gravitons.

\medskip

Thus, it follows that all HK gravitons and HH gluons on $\C^{4m}$ have a simple momentum of the form \eqref{snv}, with the underlying polarization vector of the form \eqref{snvpol1}. By separating out the projective scale inherent in the parametrization of a simple null momentum, this can be written as
\be\label{snv2}
k^{\alpha\dot\alpha}=\omega\,z^{\alpha}\,\tilde{z}^{\dot\alpha}\,,
\ee
where $\omega$ is the frequency, $z^{\alpha}$ are homogeneous coordinates on $\P^1$ and $\tilde{z}^{\dot\alpha}$ are homogeneous coordinates on $\P^{2m-1}$. Therefore, for HK/HH fields, the complexified celestial sphere $S^{4m-2}_{\C}$ is replaced by the subvariety $\P^1\times\P^{2m-1}$. In particular, there is an emergent 2-sphere $\P^1\cong S^2$, with affine holomorphic coordinate $z$ appearing as
\be\label{holsphere}
z_{\alpha}=(1,z)\,,
\ee
on the $z_0\neq 0$ patch of $\P^1$.

The inner product between simple null momenta then takes the form
\be\label{snvip}
k_i\cdot k_j=\omega_i\,\omega_j\,\la z_i\,z_j\ra\,[\tilde{z}_i\,\tilde{z}_j]=\omega_i\,\omega_j\,(z_i-z_j)\,[\tilde{z}_i\,\tilde{z}_j]\,.
\ee
This allows us to define a notion of \emph{holomorphic collinear limit} for simple null momenta in $4m$-dimensions, as:
\be\label{holcolim}
k_i\cdot k_j\to 0\,, \qquad \mbox{as } \,z_i-z_j\to0\,, \quad [\tilde{z}_i\,\tilde{z}_j]\neq 0\,.
\ee
This is the natural generalization of the usual holomorphic collinear limit when $m=1$, with the added advantage that it coincides with a holomorphic OPE limit on the Riemann sphere parametrized by $z^{\alpha}$ or $z$ for such simple null momenta.


\subsection{Conformal primary and conformally soft wavefunctions}

The Mellin transform of a momentum eigenstate with respect to its frequency defines a conformal primary wavefunction~\cite{Pasterski:2016qvg,Pasterski:2017kqt}: a solution of the massless free field equations which transforms as a conformal primary on the celestial sphere. This general statement, of course, remains true for HK/HH fields. However, as we've seen, these states have restricted kinematics, and are specified in momentum space by a frequency $\omega$ and a point in $\P^1\times\P^{2m-1}$, viewed as a subvariety of the complexified celestial sphere. As such, HK/HH conformal primaries have an emergent \emph{two-dimensional} conformal symmetry associated with the action of Lorentz transformations on the $\P^1\cong S^2$ portion of the on-shell phase space.

Let us see how this works explicitly for a HK graviton. With momentum \eqref{snv2}, the \v{C}ech and Dolbeault representatives for the HK graviton wavefunction are
\be\label{HKme1}
\begin{split}
g&=\frac{1}{2\pi\im}\,\frac{\la \iota\,\lambda\ra^3}{\omega^2\,\la\iota\, z\ra^3}\,\frac{1}{\la\lambda\,z\ra}\,\exp\!\left(\im\,\omega\,\frac{[\mu\,\tilde{z}]\,\la\iota\,z\ra}{\la\iota\,\lambda\ra}\right)\,, \\
h&=\frac{\la \iota\,\lambda\ra^3}{\omega^2\,\la\iota\, z\ra^3}\,\bar{\delta}\!\left(\la\lambda\,z\ra\right)\,\exp\!\left(\im\,\omega\,\frac{[\mu\,\tilde{z}]\,\la\iota\,z\ra}{\la\iota\,\lambda\ra}\right)\,,
\end{split}
\ee
where 
\be\label{holdelta}
\bar{\delta}\!\left(\la\lambda\,z\ra\right):=\frac{1}{2\pi\im}\,\dbar\left(\frac{1}{\la\lambda\,z\ra}\right)\,,
\ee
is the holomorphic delta function. It is straightforward to see that feeding this into the integral formulae \eqref{HKmetpert}, \eqref{HKRiem} gives rise to the expected momentum eigenstate wavefunctions:
\be\label{HKme2}
h_{\alpha\dot\alpha\beta\dot\beta}(x)=\iota_{\alpha}\,\iota_{\beta}\,\tilde{z}_{\dot\alpha}\,\tilde{z}_{\dot\beta}\,\e^{\im\,k\cdot x}\,,
\ee
where $z_{\alpha}=(1,z)$ so that $\la\iota\,z\ra=1$.

As usual, the conformal primary wavefunction in twistor space is defined by performing a Mellin transform with respect to the frequency, $\omega$. Doing this Mellin transform for the Dolbeault representative in \eqref{HKme1} gives:
\be\label{HKcp1}
h_{\Delta}:=\int_{0}^{\infty}\d \omega\,\omega^{\Delta-1}\,h=\frac{(-\im)^{-\Delta}\,\Gamma(\Delta-2)}{[\mu\,\tilde{z}]^{\Delta-2}}\,\bar{\delta}_{\Delta}(\la\lambda\,z\ra)\,,
\ee
where 
\be\label{holdelweight}
\bar{\delta}_{\Delta}(\la\lambda\,z\ra):=\left(\frac{\la\iota\,\lambda\ra}{\la\iota\,z\ra}\right)^{\Delta+1}\,\bar{\delta}(\la\lambda,z\ra)\,,
\ee
is shorthand for a holomorphic delta function of weight $\Delta$ in $\lambda_{\alpha}$, the homogeneous coordinate on twistor lines. Observe that the functional form of these twistor conformal primary representatives is independent of the spacetime dimension: \eqref{HKcp1} are the same as the twistor representatives for positive helicity conformal primary gravitons in $d=4$~\cite{Adamo:2019ipt}.

Now, due to the factor of $\Gamma(\Delta-2)$ appearing in \eqref{HKcp1}, it follows that these wavefunctions have simple poles at the usual values $\Delta=k=2,1,0,-1,\ldots$, which -- due to the Mellin transform -- correspond to the terms in a soft expansion of the wavefunction in momentum space~\cite{Guevara:2019ypd}. \emph{Conformally soft} wavefunctions on twistor space are then defined by extracting the residues of these poles:
\be\label{HKcswf1}
h_{k}^{\mathrm{soft}}=\mathrm{Res}_{\Delta=k}h_{\Delta}=\frac{\im^{-k}}{(2-k)!}\,[\mu\,\tilde{z}]^{2-k}\,\bar{\delta}_{k}(\la\lambda\,z\ra)\,.
\ee
Now, it is easy to see that these conformally soft wavefunctions can be expanded (using the multinomial theorem) as
\be\label{HKcswf2}
h_{k}^{\mathrm{soft}}=\im^{-k}\,\bar{\delta}_{k}(\la\lambda\,z\ra)\sum_{\substack{\mbf{p}\in\N_0 \\ P=2-k}}w[\mbf{p}]\,\frac{(\tilde{z}_{\dot{1}})^{p_1}\cdots(\tilde{z}_{\dot{2m}})^{p_{2m}}}{p_1!\,p_2!\cdots p_{2m}!}\,,
\ee
where $w[\mbf{p}]=(\mu^{\dot{1}})^{p_1}\cdots(\mu^{\dot{2m}})^{p_{2m}}$ are the polynomial modes on the $\C^{2m}$ fibres of $\PT$. 

The modes in the expansion \eqref{HKcswf2} are then 
\be\label{HKcsmode1}
\im^{P-2}\,\bar{\delta}_{2-P}(\la\lambda\,z\ra)\,w[\mbf{p}]\in H^{0,1}(\PT,\cO(2))\,,
\ee
which define vertex operators in the twistor sigma model of the form
\be\label{HKcsmode2}
V[\mbf{p}](z)=\im^{P-2}\,\int\frac{\D\lambda}{\la\lambda\,o\ra^2\,\la\lambda\,\iota\ra^2}\,w[\mbf{p}]\,\bar{\delta}_{2-P}(\la\lambda\,z\ra)\,.
\ee
Under the \v{C}ech-Dolbeault isomorphism, these correspond to the family of charges
\be\label{HKcscharges}
\begin{split}
Q[\mbf{p}](z)&=\frac{\im^{P-2}}{2\pi\im}\oint_{\gamma_z}\frac{w[\mbf{p}]\,\la\iota\,\lambda\ra^{3-P}}{\la\lambda\,z\ra\,\la\iota\,z\ra^{3-P}}\,\frac{\D\lambda}{\la\lambda\,o\ra^2\,\la\lambda\,\iota\ra^2} \\
&=\frac{\im^{P-2}}{2\pi\im}\oint_{\gamma_z}g[\mbf{p}](z)\,\frac{\d\lambda}{\lambda^2}\,,
\end{split}
\ee
with the second equality following on the affine patch where $\la\iota\,z\ra=1=\la\iota\,\lambda\ra$ and $\la o\,\lambda\ra=\lambda$. Here, the contour of integration is a small circle around $\la\lambda\,z\ra=0$
\be\label{chargecontour}
\gamma_z=\left\{\left|\frac{\la\lambda\,z\ra}{\la\lambda\iota\ra}\right|<\epsilon\right\}\,,
\ee
for $\epsilon$ sufficiently small enough that this is the only pole in the $\lambda$-plane which falls inside the contour.

In summary, the conformally soft HK degrees of freedom can be expanded in the modes of $\cL\mathfrak{ham}(\C^{2m})$, which in turn induce a family of charges in the twistor sigma model describing the HK sector.

\medskip

The story proceeds along similar lines for the conformal primary wavefunctions of HH gluons. In this case, the twistor Dolbeault representative for a conformal primary HH gluon is~\cite{Adamo:2019ipt}
\be\label{HHcp1}
\cA_{\Delta}=\frac{(-\im)^{\Delta-1}\,\Gamma(\Delta-1)}{[\mu\,\tilde{z}]^{\Delta-1}}\,\bar{\delta}_{\Delta-1}(\la\lambda\,z\ra)\,,
\ee
which has poles at $\Delta=k=1,0,-1,-2,\ldots$ corresponding to the conformally soft tower. The resulting conformally soft wavefunctions are
\be\label{HHcswf1}
\begin{split}
\cA_{k}^{\mathrm{soft}}&=\mathrm{Res}_{\Delta=k}\cA_{\Delta}=\frac{\im^{1-k}}{(1-k)!}\,[\mu\,\tilde{z}]^{1-k}\,\bar{\delta}_{k-1}(\la\lambda\,z\ra) \\
&=\im^{1-k}\,\bar{\delta}_{k-1}(\la\lambda\,z\ra) \sum_{\substack{\mbf{p}\in\N_0 \\ P=1-k}}w[\mbf{p}]\,\frac{(\tilde{z}_{\dot{1}})^{p_1}\cdots(\tilde{z}_{\dot{2m}})^{p_{2m}}}{p_1!\,p_2!\cdots p_{2m}!}\,,
\end{split}
\ee
where we have expanded the wavefunction in modes in the second line using the multinomial theorem.

The modes
\be\label{HHcsmode1}
\im^{P}\,\bar{\delta}_{-P}(\la\lambda\,z\ra)\,w[\mbf{p}]\in H^{0,1}(\PT,\cO)\,,
\ee
then define vertex operators 
\be\label{HHcsmode2}
V^{\sa}[\mbf{p}](z)=\im^{P}\int j^{\sa}\wedge \bar{\delta}_{-P}(\la\lambda\,z\ra)\,w[\mbf{p}]\,,
\ee
and the corresponding charges
\be\label{HHcscharges}
\begin{split}
Q^{\sa}[\mbf{p}](z)&=\frac{\im^P}{2\pi\im}\oint_{\gamma_z}\frac{w[\mbf{p}]\,\la\iota\,\lambda\ra^{1-P}}{\la\lambda\,z\ra\,\la\iota\,z\ra^{1-P}}\,j^{\sa} \\
&=\frac{\im^P}{2\pi\im}\oint_{\gamma_z} \bar{\rho}\,S^{\sa}[\mbf{p}](z)\,\rho\,,
\end{split}
\ee
in the twistor sigma model. So, we see that the conformally soft HH degrees of freedom can be expanded in modes of $\cL\mathfrak{g}[\C^{2m}]$, inducing a family of charges in the twistor sigma model.


\subsection{Soft HK graviton symmetries}

It is now possible to compute the action of a charge \eqref{HKcscharges} corresponding to the modes of a conformally soft HK graviton on a \emph{hard} HK graviton. The action of these charges is governed by the semiclassical OPE \eqref{muOPE} of the twistor sigma model, which gives rise to
\begin{multline}\label{sHKOPE1}
Q[\mbf{p}](z)\,h_{\Delta}(Z;z',\tilde{z})\sim\frac{\im^{P-2}}{2\pi\im}\oint\frac{\la\lambda\,\iota\ra\,\la\lambda\,o\ra}{\la\lambda'\,\lambda\ra\,\la\iota\,\lambda'\ra\,\la o\,\lambda'\ra}\,\frac{\partial w[\mbf{p}]}{\partial\mu_{\dot\alpha}}\,\frac{\partial h_{\Delta}}{\partial\mu^{\dot\alpha}}\,\frac{\D\lambda'\,\la\iota\,\lambda'\ra^{3-P}}{\la\lambda'\,z\ra\,\la\iota\,z\ra^{3-P}} \\
=\frac{\im^{P-2}}{\la z\,\lambda\ra}\,\frac{\la\iota\,\lambda\ra^{3-P}}{\la\iota\,z\ra^{3-P}}\,\frac{\partial w[\mbf{p}]}{\partial\mu_{\dot\alpha}}\,\frac{\partial h_{\Delta}}{\partial\mu^{\dot\alpha}}(Z;z',\tilde{z})\,,
\end{multline}
with the second line following after deforming the contour of integration to surround the pole $\la\lambda\,\lambda'\ra=0$ in the first line. 

Now, recall that in the hard HK graviton wavefunction \eqref{HKcp1} there is the holomorphic delta function $\bar{\delta}_{\Delta}(\la\lambda\,z'\ra)$. On the support of this delta function, we can replace
\be\label{Hdeltreplace}
\la z\,\lambda\ra\rightarrow \la z\,z'\ra\,\frac{\la\iota\,\lambda\ra}{\la\iota\,z'\ra}\,,
\ee
which gives
\be\label{sHKOPE2}
\begin{split}
Q[\mbf{p}](z)\,h_{\Delta}(Z;z',\tilde{z})&\sim\frac{\im^{P-2}}{\la z\,z'\ra}\,\frac{\la\iota\,\lambda\ra^{2-P}\,\la\iota\,z'\ra}{\la\iota\,z\ra^{3-P}}\,\left\{w[\mbf{p}],\,h_{\Delta}(Z;z',\tilde{z})\right\} \\
&\sim \frac{\im^{P-2}}{z-z'}\,\frac{\{w[\mbf{p}],\,h_{\Delta}(Z;z',\tilde{z})\}}{\la\iota\,\lambda\ra^{P-2}}\,.
\end{split}
\ee
Here, the second line is the result evaluated in the patch where $\la\iota\,z\ra=1=\la\iota\,z'\ra$. This shows that the OPE between the conformally soft charges and the HK graviton wavefunction is equivalent to a \emph{celestial OPE}; namely, \eqref{sHKOPE2} takes the form of a holomorphic collinear singularity on the Riemann sphere parametrized by the SL$(2,\C)$ components of the HK graviton momenta. In particular, although the full (complexified) celestial sphere is $(4m-2)$-complex-dimensional, the HK kinematics isolate a holomorphic collinear singularity on a one-dimensional complex subvariety of this space.

At this stage, \eqref{sHKOPE2} gives the action of HK symmetries on twistor space, but this can be rewritten to give the action directly on the celestial variety $\P^1\times\P^{2m-1}$ by computing the Poisson bracket explicitly. To do this, we work on the coordinate patch where $\tilde{z}_{\dot{1}}\neq 0$, identifying
\be\label{aholpatch1}
\tilde{z}_{\dot\alpha}=(1,\tilde{z}_{2},\ldots,\tilde{z}_{2m})=(1,\tilde{z}_{a})\,,
\ee
for $\tilde{z}_{a}$, $a=2,\ldots,2m$, the affine coordinates on this patch. A straightforward, albeit tedious, calculation shows that
\begin{multline}\label{sHKOPE3}
\left\{w[\mbf{p}],\,[\mu\,\tilde{z}]^{2-\Delta}\right\}=\frac{(-1)^{P-p_1+1}}{(\Delta-3)_{P-2}}\left[p_1\,\tilde{z}_{m+1}\left(\prod_{r=0}^{p_1-2}\tilde{\Upsilon}+\Delta-2-r\right)\prod_{b=2}^{2m}\tilde{\partial}_b^{p_b} \right. \\
-p_{1+m}\left(\prod_{r=0}^{p_1-1}\tilde{\Upsilon}+\Delta-2-r\right) \tilde{\partial}_{1+m}^{p_{1+m}-1}\prod_{b\neq1,1+m}\tilde{\partial}_{b}^{p_b}  \\
+\sum_{i=2}^{m}\left(p_i\,\tilde{z}_{i+m}\left(\prod_{r=0}^{p_1-1}\tilde{\Upsilon}+\Delta-2-r\right)\tilde{\partial}_i^{p_i-1}\prod_{b\neq1,i}\tilde{\partial}_{b}^{p_b}\right. \\
\left.\left.-p_{i+m}\,\tilde{z}_{i}\left(\prod_{r=0}^{p_1-1}\tilde{\Upsilon}+\Delta-2-r\right)\tilde{\partial}_{i+m}^{p_{i+m}-1}\prod_{b\neq1,i+m}\tilde{\partial}_{b}^{p_b}\right)\right][\mu\,\tilde{z}]^{P-\Delta}\,,
\end{multline}
where $(a)_b$ is the falling factorial, $\tilde{\partial}_{a}\equiv\frac{\partial}{\partial\tilde{z}_a}$ and $\tilde{\Upsilon}$ is the differential operator
\be\label{tildeups}
\tilde{\Upsilon}:=\tilde{z}_{a}\,\frac{\partial}{\partial\tilde{z}_a}\,.
\ee
Combining this with the pre-factors appearing in \eqref{sHKOPE2} gives
\begin{multline}\label{sHKOPE4}
Q[\mbf{p}](z)\,h_{\Delta}(z',\tilde{z})\sim\frac{-(-1)^{P-p_1}}{z-z'}\left[p_1\,\tilde{z}_{m+1}\left(\prod_{r=0}^{p_1-2}\tilde{\Upsilon}+\Delta-2-r\right)\prod_{b=2}^{2m}\tilde{\partial}_b^{p_b} \right. \\
-p_{1+m}\left(\prod_{r=0}^{p_1-1}\tilde{\Upsilon}+\Delta-2-r\right) \tilde{\partial}_{1+m}^{p_{1+m}-1}\prod_{b\neq1,1+m}\tilde{\partial}_{b}^{p_b}  \\
+\sum_{i=2}^{m}\left(p_i\,\tilde{z}_{i+m}\left(\prod_{r=0}^{p_1-1}\tilde{\Upsilon}+\Delta-2-r\right)\tilde{\partial}_i^{p_i-1}\prod_{b\neq1,i}\tilde{\partial}_{b}^{p_b}\right. \\
\left.\left.-p_{i+m}\,\tilde{z}_{i}\left(\prod_{r=0}^{p_1-1}\tilde{\Upsilon}+\Delta-2-r\right)\tilde{\partial}_{i+m}^{p_{i+m}-1}\prod_{b\neq1,i+m}\tilde{\partial}_{b}^{p_b}\right)\right]\,h_{\Delta-P+2}(z,\tilde{z})\,,
\end{multline}
for the action of the HK symmetries on generic (hard) conformal primary HK gravitons. This expression is now entirely in momentum space, and can be interpreted as a celestial OPE for the action of the HK charges.

By expanding out the products appearing in \eqref{sHKOPE4}, it is possible to partially resum this expression, obtaining the slightly more compact formula
\begin{multline}\label{sHKOPE5}
Q[\mbf{p}](z)\,h_{\Delta}(z',\tilde{z})\sim \\
\frac{-(-1)^{P-p_1}}{z-z'}\left[\sum_{L=0}^{p_1-1}\frac{\Gamma(\Delta-1)}{\Gamma(\Delta-p_1+L)}\,\binom{p_1-1}{L}\sum_{\substack{\ell_2,\ldots,\ell_{2m}\geq0 \\ \ell_{2}+\cdots+\ell_{2m}=L}}\frac{L!\,p_{1}\,\tilde{z}_{m+1}}{\ell_2!\cdots\ell_{2m}!}\,\prod_{b=2}^{2m}\tilde{z}_b^{\ell_b}\,\tilde{\partial}_b^{\ell_b+p_b}\right. \\
-\sum_{L=0}^{p_1}\frac{\Gamma(\Delta-1)}{\Gamma(\Delta-1-p_1+L)}\,\binom{p_1}{L}\sum_{\substack{\ell_2,\ldots,\ell_{2m}\geq0 \\ \ell_{2}+\cdots+\ell_{2m}=L}}\frac{L!\,p_{m+1}}{\ell_2!\cdots\ell_{2m}!}\,\tilde{z}_{m+1}^{\ell_{m+1}}\,\tilde{\partial}_{m+1}^{p_{m+1}+\ell_{m+1}-1}\prod_{b\neq m+1}^{2m}\tilde{z}_b^{\ell_b}\,\tilde{\partial}_b^{\ell_b+p_b} \\
+\sum_{i=2}^{m}\sum_{L=0}^{p_1}\frac{\Gamma(\Delta-1)}{\Gamma-1-p_1+L)}\,\binom{p_1}{L}\sum_{\substack{\ell_2,\ldots,\ell_{2m}\geq0 \\ \ell_{2}+\cdots+\ell_{2m}=L}}\frac{L!}{\ell_2!\cdots\ell_{2m}!}\left(p_i\,\tilde{z}_{i+m}\,\tilde{z}_i^{\ell_i}\,\tilde{\partial}_{i}^{\ell_i+p_i-1}\prod_{b\neq i}\tilde{z}_b^{\ell_b}\,\tilde{\partial}_b^{\ell_b+p_b}  \right.\\
-\left.\left. p_{m+i}\,\tilde{z}_i\,\tilde{z}_{i+m}^{\ell_{i+m}}\,\tilde{\partial}_{i+m}^{\ell_{i+m}+p_{i+m}-1}\prod_{b\neq i+m}\tilde{z}_b^{\ell_b}\,\tilde{\partial}_b^{\ell_b+p_b}\right)\right]h_{\Delta-P+2}(z,\tilde{z})\,.
\end{multline}
It is easy to see that when $m=1$, this expression reduces to formulae for the action of a $\cL\mathfrak{ham}(\C^2)$ charge on a positive helicity hard graviton (cf., \cite{Himwich:2021dau,Jiang:2021ovh,Adamo:2021lrv}).


\subsection{Soft HH gluon symmetries}

It is also possible to compute the action of HH symmetry charges on HH conformal primary gluons; this calculation follows similar lines to the gravitational one above, with some minor distinctions. Firstly, because the dynamics of the HH sector are encoded in the Kac-Moody currents \eqref{KMcurr} and their OPEs \eqref{jjOPE}, one must consider the charges acting on the \emph{vertex operators} \eqref{gluVO}, rather than the bare wavefunction \eqref{HHcp1} in isolation. 

This means that one wishes to consider the semiclassical action of $Q^{\sa}[\mbf{p}](z)$ on the conformal primary vertex operator
\be\label{HHcp2}
V_{\Delta}^{\msf{b}}(z',\tilde{z})=(-\im)^{\Delta-1}\,\Gamma(\Delta-1)\int j^{\msf{b}}\wedge\frac{\bar{\delta}_{\Delta-1}(\la\lambda\,z'\ra)}{[\mu\,\tilde{z}]^{\Delta-1}}\,.
\ee
Using the OPE \eqref{jjOPE}, it is straightforward to show that this gives
\be\label{sHHOPE1}
Q^{\sa}[\mbf{p}](z)\,V^{\msf{b}}_{\Delta}(z')\sim\frac{(-1)^{P+1}\,f^{\msf{abc}}}{z-z'}\,(-\im)^{\Delta-P-1}\,\Gamma(\Delta-1)\int j^{\msf{c}}\wedge \bar{\delta}_{\Delta-P-1}(\la\lambda\,z'\ra)\,w[\mbf{p}]\,[\mu\,\tilde{z}]^{1-\Delta}\,,
\ee
after performing the contour integral and using the support \eqref{Hdeltreplace} of the holomorphic delta function. This can in turn be rewritten on the affine patch \eqref{aholpatch1} as:
\be\label{sHHOPE2}
Q^{\sa}[\mbf{p}](z)\,V^{\msf{b}}_{\Delta}(z')\sim\frac{f^{\msf{abc}}}{z-z'}\,\left(\prod_{r=0}^{p_1-1}\tilde{\Upsilon}+\Delta-2-r\right)\prod_{b=2}^{2m}\tilde{\partial}_b^{p_b}\,V^{\msf{c}}_{\Delta-P}(z,\tilde{z})\,,
\ee
giving the celestial action of the HH charges on hard HH modes, as desired. Expanding the product over the differential operators, one obtains the equivalent resummed formula
\begin{multline}\label{sHHOPE3}
Q^{\sa}[\mbf{p}](z)\,V^{\msf{b}}_{\Delta}(z')\sim\frac{f^{\msf{abc}}}{z-z'}\sum_{L=0}^{p_1}\frac{\Gamma(\Delta-1)}{\Gamma(\Delta-1-p_1+L)}\,\binom{p_1}{L} \\
\times\,\sum_{\substack{\ell_2,\ldots,\ell_{2m}\geq0 \\ \ell_{2}+\cdots+\ell_{2m}=L}}\frac{L!}{\ell_2!\cdots\ell_{2m}!}\prod_{b=2}^{2m}\tilde{z}_b^{\ell_b}\,\tilde{\partial}_b^{\ell_b+p_b}\,V^{\msf{c}}_{\Delta-P}(z,\tilde{z})\,,
\end{multline}
for the celestial action on a hard HH gluon. When $m=1$, this reduces to the known expressions for the action of $\cL\mathfrak{g}[\C^2]$ on positive helicity gluons (cf., \cite{Himwich:2021dau,Jiang:2021ovh,Adamo:2021zpw}).


\section{Conclusion}
\label{sec 6}

In this paper, we introduced and described twistorial chiral algebras in higher dimensions as either the algebra of Poisson diffeomorphisms on the fibers of twistor space or the algebra of gauge symmetries of the Ward bundle. The modes of conformally soft hyperholomorphic gluons or hyperk\"ahler gravitons organize themselves into $\mathcal{L}\mathfrak{g}[\mathbb{C}^{2m}]$ and $\mathcal{L}\mathfrak{ham}(\mathbb{C}^{2m})$, respectively. These algebras are realised dynamically through charges in twistor sigma models, which enables the identification of the OPE between these charges and hard degrees of freedom with a holomorphic celestial OPE, thanks to the restricted kinematics of the HH/HK sectors. There are several open questions and research directions raised by this work, which we comment on briefly here. 

In four dimensions, the singular part of celestial OPEs can be derived from the splitting functions appearing in holomorphic collinear limits of the MHV sector in gauge theory and gravity~\cite{Fan:2019emx,Guevara:2019ypd}. It is natural then to seek equivalent `MHV-like' expressions in $4m$-dimensions. This `MHV-like' configuration would, at tree-level, correspond to a scattering amplitude with two non-HH/HK and an \emph{arbitrary} number of HH/HK gluons/gravitons, and we expect remarkably compact formulae for these amplitudes which generalise the Parke-Taylor~\cite{Parke:1986gb} and Hodges~\cite{Hodges:2012ym} formulae in four-dimensions. In the gauge theory case, we will report on results confirming these speculations in the near future. 

It is also interesting to contrast the twistor construction for $m>1$ with the well-studied four-dimensional case. In 4d, twistor space admits a projection to the complexified null conformal boundary of Minkowski spacetime, $\PT\to\scri_{\C}$, where $\scri_{\C}\cong\C\times\P^1$. Under this projection, the twistor lines are identified with the celestial sphere, with a given twistor line identified with the cut of $\scri_{\C}$ defined by the intersection with a lightcone whose apex is the corresponding spacetime point~\cite{Eastwood:1981jy}. This identification also extends to the self-dual degrees of freedom: the cohomology classes (\v{C}ech or Dolbeault) on $\PT$ encoding self-dual perturbations can be mapped to radiative data for the positive helicity gluons or gravitons, so that the Kirchoff-d'Adh\'emar integral formula (which constructs the radiative field from its characteristic data at $\scri$~\cite{Penrose:1980yx}) is simply a gauge-fixed version of the Penrose transform integral formula (cf., \cite{Mason:1986,Adamo:2021dfg,Adamo:2021lrv}). Non-linearly, this allows the radiative data of any asymptotically flat 4d spacetime to be used to construct an `asymptotic' twistor space via Newman's $\cH$-space construction~\cite{Newman:1976gc,Hansen:1978jz,Ko:1981}.

In $4m$-dimensions for $m>1$, the relationship between the twistor data and radiative data in spacetime is less clear. Here, the twistor space is defined with reference to the HK structure of spacetime, which is a (complexification of a) Riemannian, rather than Lorentzian, structure. Consequently, there is no obvious projection from $\PT$ to $\scri_{\C}\cong\C\times S^{4m-2}$, and twistor lines $X\cong\P^1$ cannot be identified with $S^{4m-2}$ cuts of $\scri_{\C}$. Nevertheless, one still expects a higher-dimensional analogy of the Kirchoff-d'Adh\'emar integral formula to hold for general radiative fields, and when these fields are HH/HK this description must be equivalent to the one provided by Penrose transform. It would be interesting to understand this correspondence, which must rely on the constrained kinematics of the HH/HK sectors to pick out a $S^2$ within the higher-dimensional celestial sphere that is identified with the twistor lines. This could have a nice manifestation in the Geroch-Held-Penrose formalism~\cite{Geroch:1973am}, a higher-dimensional generalisation of the Newman-Penrose formalism~\cite{Newman:1961qr}, which explicitly parametrizes the radiative degrees of freedom.

Finally, recall that twistorial chiral algebras provide the boundary theories for all known top-down constructions of celestial holography~\cite{Costello:2022jpg,Costello:2023hmi,Bittleston:2024efo}, which are realised through twisted holography for topological string theories in twistor space. While these examples are all in four-dimensions, it would be interesting to find such constructions for higher-dimensional hyperk\"ahler manifolds by considering string theories with the HK twistor space as a target.

\acknowledgments

TA thanks Glenn Barnich, Guillaume Bossard, Laura Donnay, Yorgo Pano, Marios Petropoulos and Andrea Puhm for interesting conversations. TA is supported by a Royal Society University Research Fellowship, the Simons Collaboration on Celestial Holography CH-00001550-11, the ERC Consolidator/UKRI Frontier grant TwistorQFT EP/Z000157/1 and the STFC consolidated grant ST/X000494/1. IS is supported by an STFC studentship.

\bibliographystyle{JHEP}
\bibliography{HyperH}

\end{document}